\documentclass[10pt,journal,compsoc]{IEEEtran}
\usepackage{amsmath}
\usepackage{graphicx}       
\usepackage{booktabs}
\usepackage{tabularx}
\usepackage{multirow}
\usepackage{cases}
\usepackage{mathrsfs}
\usepackage{amssymb}
\usepackage{algorithm}
\usepackage{algorithmic}
\usepackage{xcolor}
\usepackage{siunitx}
\usepackage{multirow}
\usepackage{diagbox}
\usepackage{enumitem}
\usepackage[table]{xcolor}

\ifCLASSOPTIONcompsoc
  \usepackage[nocompress]{cite}
\else
  \usepackage{cite}
\fi

\ifCLASSINFOpdf
\else
\fi

\begin{document}
%
\title{Spreading Dynamics on Higher-Order Networks: A Structure–Mechanism–Behavior Perspective with Applications}
%
%
%
%

\author{Ming Xie,~\IEEEmembership{Student Member,~IEEE}, Shibo~He,~\IEEEmembership{Senior Member,~IEEE}, Qihao~Huang, Yuzhen Fan~and Jiming~Chen,~\IEEEmembership{Fellow,~IEEE}
\IEEEcompsocitemizethanks{\IEEEcompsocthanksitem This version corresponds to the manuscript originally submitted to a peer-reviewed journal on April 21, 2026.
\IEEEcompsocthanksitem Ming Xie, Shibo He, Qihao Huang, Yuzhen Fan and Jiming Chen are with the State Key Lab. of Industrial Control Technology, Zhejiang University, Hangzhou 310027, China
(e-mail: mingxie@zju.edu.cn; s18he@zju.edu.cn; 22332052@zju.edu.cn; yzfan@zju.edu.cn; cjm@zju.edu.cn). }
}

\IEEEtitleabstractindextext{%
\begin{abstract}
Spreading in real-world networks, including modern communication and networked systems, often unfolds through group interactions rather than isolated pairwise contacts. Higher-order network representations such as simplicial complexes, hypergraphs, and their temporal, multilayer, and directed extensions provide a principled way to encode group interactions. Unfortunately, existing studies on spreading over these substrates remain fragmented and largely model-centric, offering limited insight into how higher-order representations and contagion mechanisms jointly shape emergent dynamics. This review addresses this gap by introducing a structure–mechanism–behavior three-level perspective that provides a systematic mapping from higher-order network structures and spreading mechanisms to collective behaviors. Within this framework, we compare how different higher-order network representations encode group interactions, introduce a taxonomy of contagion mechanisms, and analyze their structural carrying capacity, defined as the ability of a given representation to support specific contagion mechanisms and the associated dynamical behaviors.
We further map specific structure–mechanism combinations to three hallmark emergent behaviors, thereby offering a unified cross-level perspective beyond existing model-centric surveys. These insights are particularly relevant for communication systems involving group interactions, such as social information spreading, cooperative protocols, and distributed networked processes.
Lastly, leveraging this three-level perspective, we offer a roadmap for guiding and controlling dynamics on higher-order networks, while discussing future directions in representation choice, mechanism design, and data-driven inference.
\end{abstract}

}

\maketitle

\IEEEdisplaynontitleabstractindextext

%
\IEEEpeerreviewmaketitle

\IEEEraisesectionheading{\section{Introduction}\label{sec:introduction}}
\IEEEPARstart{I}{n} many real-world systems, particularly social environments and information-sharing networks, contagion unfolds not through isolated one-to-one contacts but through collective interactions~\cite{kiss2023insights,battiston2020networks,majhi2022dynamics}. 
Behavioral change is often negotiated in groups, where people adopt novel technologies in project teams, adjust health practices such as mask wearing or vaccination within families, classrooms, or offices, and align opinions during in-person meetings rather than through a single persuasive neighbor~\cite{centola2010spread,o2015mathematical}. In online social platforms, e.g., WhatsApp and Telegram, information and misinformation typically spread via group chats, comment threads, and community channels, where several individuals are exposed at once and can observe each other’s responses in real time~\cite{iacopini2024temporal,alvarez2021evolutionary}.
Similar patterns arise beyond social systems: biochemical reactions require the joint presence of multiple molecules~\cite{mayfield2017higher,bairey2016high}, and collective animal motion emerges from multi-body coordination~\cite{macgregor2020information}. 
Across these domains, the effective “unit of interaction” is therefore a group configuration rather than an individual link in a network, with important implications for how \emph{contagion dynamics} (also termed \emph{spreading dynamics} in modeling) are modeled and understood.

Group interactions are not merely a modeling refinement, but play a crucial role in reshaping how spreading unfolds~\cite{battiston2021physics}. 
Spreading influence that is weak in isolated one-to-one contacts can become significant when multiple individuals act simultaneously in a group interaction~\cite{centola2010spread,centola2007complex}. This is particularly evident in small, tightly connected groups such as households, dormitories, or work teams, where repeated interactions reinforce behavioral alignment~\cite{miller2009spread,salathe2010dynamics}.
Empirical studies show that such group effects can generate explosive contagion, abrupt outbreaks, and long latent periods followed by sudden cascades, patterns that are difficult to reconcile with models treating all interactions as independent pairwise contacts~\cite{iacopini2019simplicial,de2020social,kilic2022simplicial}. 
Revealing how group interactions feed into spreading dynamics is therefore crucial for interpreting data, anticipating systemic risks, and designing interventions in realistic settings for communication and networked systems.

Although classical models of spreading dynamics represent networked systems as graphs~\cite{allen1994some} and their temporal~\cite{holme2016temporal} or multilayer~\cite{salehi2015spreading, de2016physics} extensions, they are fundamentally based on pairwise interactions, which limits their ability to capture group-driven spreading dynamics observed in many real-world systems. 
This limitation has motivated the development of higher-order network representations, such as simplicial complexes and hypergraphs, where interactions among three or more nodes are treated as fundamental building blocks rather than being reconstructed from edges~\cite{iacopini2019simplicial,st2024nonlinear}. 

\begin{figure*}[t]
  \centering
  \includegraphics[width=.9\linewidth]{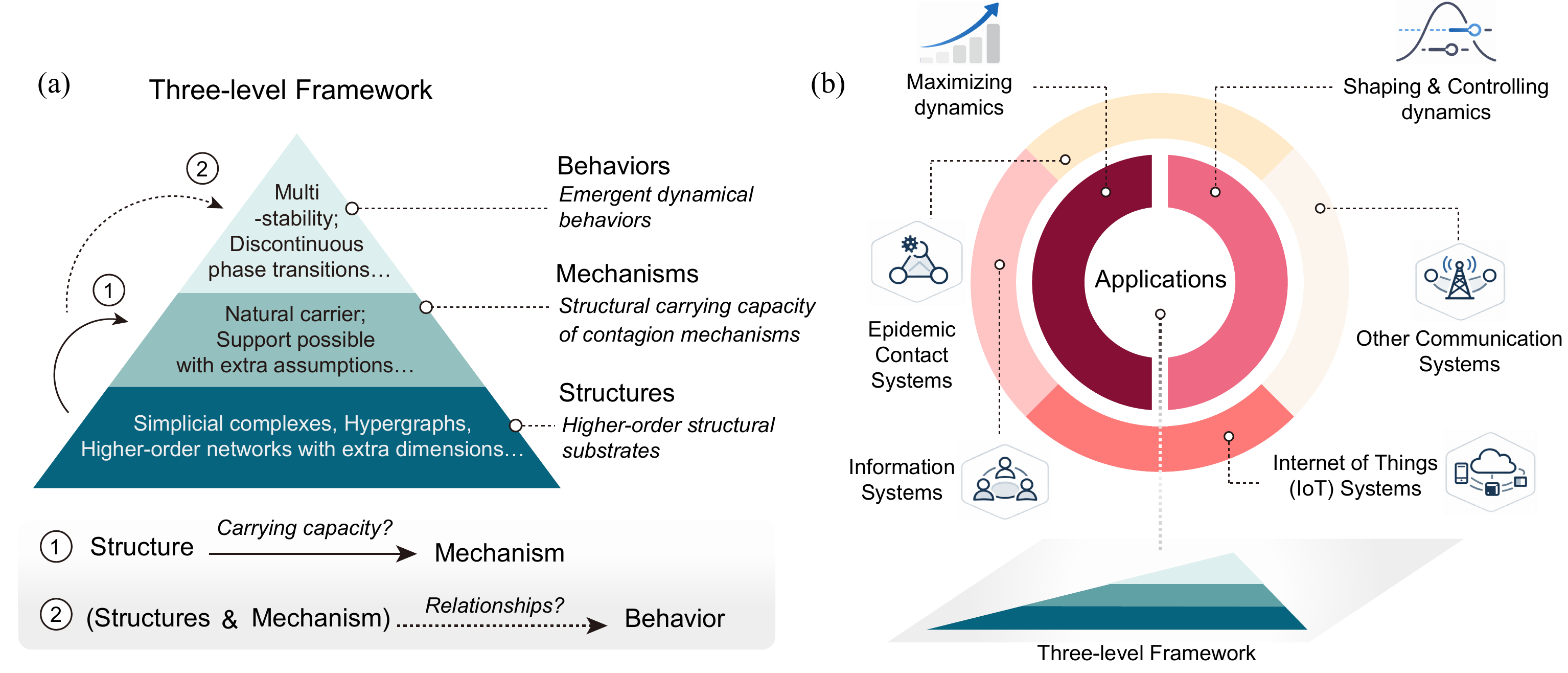}
  \caption{Schematic overview of the review. (a) The illustration of the three-level framework. We (i) identify higher-order network representations capable of encoding group-level interactions, (ii) analyze how different structures support or limit distinct classes of contagion mechanisms, and (iii) map structure–mechanism combinations to emergent dynamical behaviors such as multistability, discontinuous or hybrid phase transitions, and propagation inertia. (b) Applications for guiding and controlling
dynamics on higher-order networks by adopting a structure–mechanism–behavior three-level
perspective in different domains, including epidemic contact systems, information systems, Internet of Things (IoT) systems and other communication systems.}
  \label{figure1}
\end{figure*}

In this review, we focus on spreading dynamics defined directly on higher-order network structures, where the state of a node can depend on the joint configuration of groups rather than only on pairwise neighbors. Such processes can be captured using simplices, hyperedges, or time-stamped group interactions, providing a unified framework for modeling group-driven spreading phenomena. Research on spreading dynamics in higher-order settings has developed rapidly across statistical physics, applied mathematics, and computational and social sciences. Early studies extended classical epidemic and social spreading models from graphs to simplicial complexes and hypergraphs, revealing how explicit group interactions can alter dynamical thresholds and phase transitions~\cite{iacopini2019simplicial,jhun2019simplicial,de2020social}. Subsequent work introduced mean-field, spectral, and operator-based descriptions of higher-order processes, alongside increasing adoption in empirical studies of collective behavior~\cite{landry2020effect,ferraz2021phase}. These works primarily established the foundational understanding of higher-order spreading dynamics. Nevertheless, the field remains fragmented across disciplines, with limited shared data, computational tools, and unified modeling frameworks, leading to a dispersed set of results across specialized studies.

Interest in higher-order networks has intensified in recent years, with spreading dynamics on these structures becoming a particularly active research focus. A growing body of work has revealed multistability~\cite{ferraz2023multistability}, discontinuous and hybrid phase transitions~\cite{matamalas2020abrupt}, and pronounced propagation inertia~\cite{lv2024delay} that do not appear (or appear only in much more constrained form) on purely pairwise networks.  These behaviors, in particular, highlight distinct aspects of higher-order contagion: multistability and coexistence reveal the system’s sensitivity to initial conditions and the possibility of multiple stable outcomes; discontinuous transitions demonstrate abrupt collective responses that are absent in pairwise models; and propagation inertia captures temporal and memory effects that can significantly delay or suppress outbreaks.
At the same time, higher-order tools have begun to be used in a wide range of applications across various domains, from social and information systems to biological, technological, and infrastructural networks~\cite{liu2024higher,santoro2024higher,chen2023learning}.

Despite these advances in higher-order networked systems, both the role of higher-order structures in supporting distinct contagion mechanisms and the coupled structure–mechanism effects on emergent behaviors remain fragmented. Existing studies often focus on a few canonical higher-order representations, such as simplicial complexes and hypergraphs, and analyze particular contagion rules on each of them~\cite{suo2018information,xie2023vital}. Several reviews have synthesized findings on higher-order interactions and contagion dynamics. For example,~\cite{majhi2022dynamics} provides a broad overview of dynamical phenomena on higher-order networks, including synchronization, consensus, cooperation, and contagion, and highlight general open challenges. ~\cite{ferraz2024contagion} focuses specifically on contagion dynamics, proposing a unified formalism to encompass various epidemic and social contagion models and outlining a research agenda. More recently,~\cite{wang2024epidemic} presents a comprehensive survey of epidemic-based spreading on higher-order networks, summarizing modeling frameworks, analytical approaches, and key dynamical results. These works have been instrumental in defining the field’s contours.
However, these existing reviews generally remain centered on specific model families or process formulations, focusing on particular representations or dynamical processes in isolation. As a result, they do not explicitly disentangle the relationship between higher-order structural representations and contagion mechanisms, nor provide a systematic taxonomy of spreading mechanisms across different modeling frameworks, nor characterize how the interplay between structure and dynamics gives rise to distinct emergent behaviors across models and representations.
Consequently, several key questions remain open:
\begin{itemize}
    \item How can spreading dynamics on higher-order networks be organized into a mechanism-driven taxonomy across different interaction structures and modeling frameworks?
    \item Which spreading mechanisms are genuinely distinct, and which can be meaningfully carried by different higher-order structures?
    \item How do combinations of structures and mechanisms give rise to characteristic emergent behaviors such as multistability, hybrid phase transitions, and propagation inertia?
    \item How can these insights guide the selection of structural representations and contagion models in practical applications to anticipate or control spreading dynamics?
\end{itemize}
Addressing these issues requires understanding which structures can sustain which mechanisms and how they give rise to distinct emergent behaviors. Such a systematic mapping of structure–mechanism–behavior relationships not only advances the theoretical analysis of complex contagion, but also supports practical applications, including data-driven inference of infection kernels and thresholds, anticipating system-level risks, and guiding interventions in social, biological, and technological intelligent networked systems with higher-order interactions.

To address these questions, this review adopts a \emph{structure–mechanism–behavior perspective} with three main goals, see Fig.~\ref{figure1}. First, we clarify the space of higher-order network structures relevant for contagion, including simplicial complexes, hypergraphs, temporal higher-order networks, and multiplex higher-order networks (Section~\ref{preliminaries}). Second, we introduce a taxonomy of contagion mechanisms (Section~\ref{mechanisms}), including synergistic reinforcement, competitive inhibition, memory-aware dynamics, and threshold-activated processes, and analyze how these mechanisms are supported by different higher-order network representations, a mapping we refer to as \emph{structural carrying capacity}. Third, we map combinations of structures and mechanisms to emergent dynamical behaviors (i.e., multistability and
coexistence, discontinuous/hybrid phase transitions, and prop-
agation inertia), capturing the distinctive ways higher-order interactions reshape contagion dynamics (Section~\ref{behaviors}). Additionally, we provide applications of this mapping and offer design principles to guide and control spreading dynamics on higher-order networks (Section~\ref{applications}). Finally, we concludes this work (Section~\ref{conclusion}).


\begin{figure}[t]
  \centering
  \includegraphics[width=\linewidth]{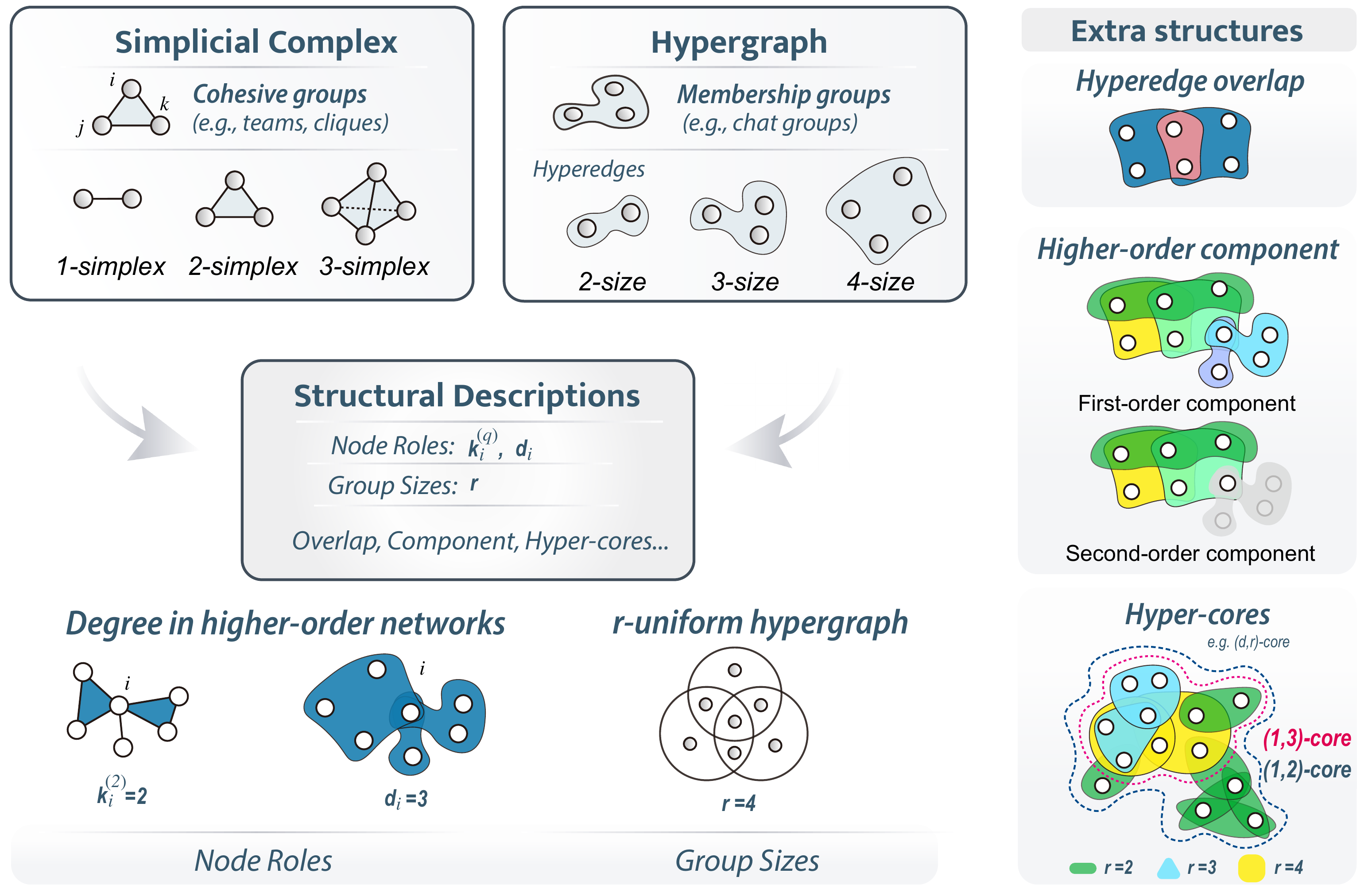}
\caption{Structural representations and descriptors of higher-order networks. The figure illustrates simplicial complexes with representative $q$-simplices ($q=1,2,3$) and hypergraphs with hyperedges of different sizes (e.g., 2-, 3-, and 4-body interactions). It further highlights key structural descriptions, including node-level roles (e.g., higher-order degree $k_i$ in a simplicial complex $\mathcal{K}$ composed of 1- and 2-simplices), group sizes (e.g., a 4-uniform hypergraph), and extra mesoscopic group-based structures relevant to contagion dynamics, such as hyperedge overlap, higher-order components (including first- and second-order components~\cite{kim2024higher}), and hyper-cores~\cite{mancastroppa2023hyper}.}
  \label{figure_model}
\end{figure}

\begin{figure*}[t]
  \centering
  \includegraphics[width=\linewidth]{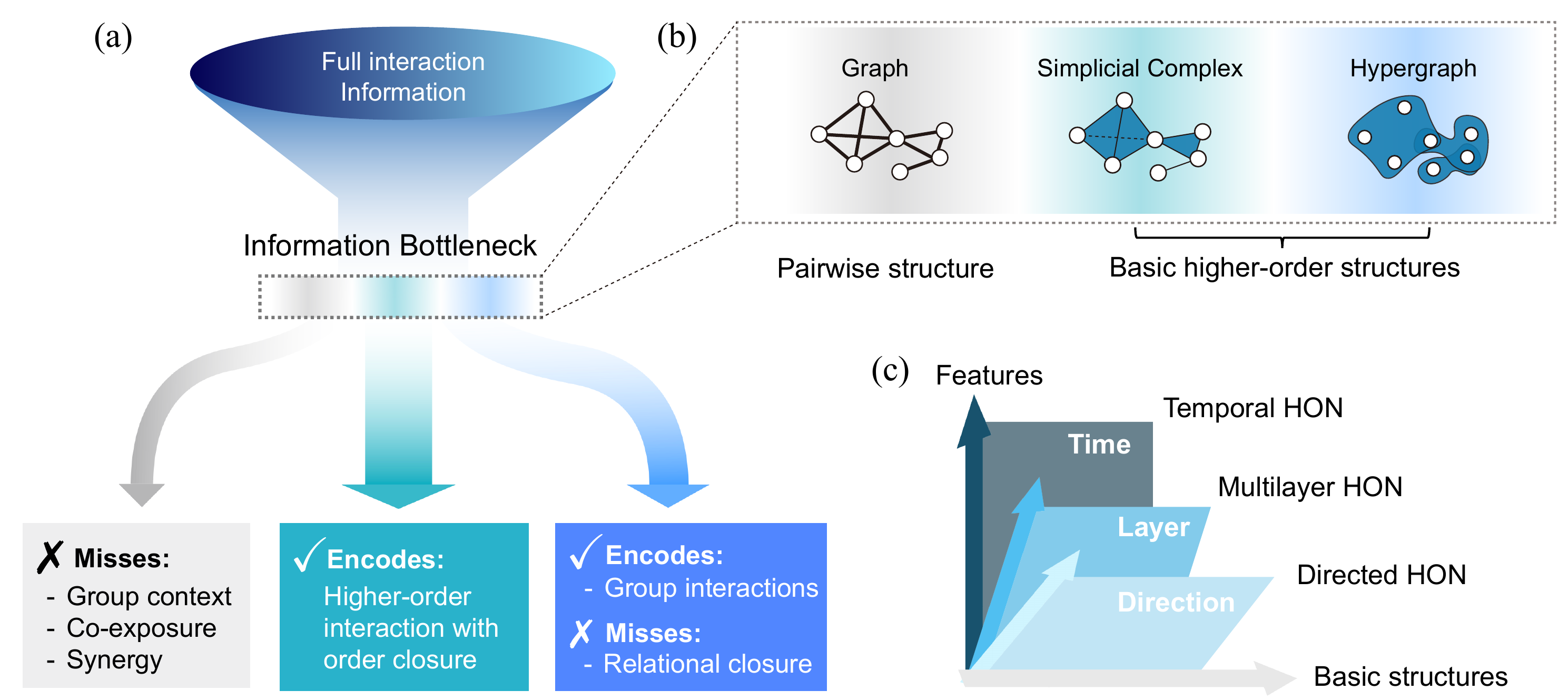}
  \caption{Basic higher-order network structures and their extensions. (a) The bottleneck illustrates different structural “channels” for graphs, simplicial complexes, and hypergraphs. The full interaction information is lost differently across channels, with graphs losing the most information compared to higher-order structures.
(b) Canonical representations of networks: a graph, a simplicial complex and a hypergraph. The latter two examples highlight the basic higher-order structures beyond simple graphs.
(c) Enriching these basic structures with temporal, multilayer, or directional features yields temporal, multilayer, and directed higher-order networks (HONs).}
  \label{figure2}
\end{figure*}

\section{Descriptions of structures}\label{preliminaries}
In this section, we introduce basic higher-order network representations, focusing on simplicial complexes and hypergraphs (see Fig.~\ref{figure_model}), and outline their extensions obtained by incorporating time, layer, and direction. Here, we do not aim to review higher-order structures in full generality, but to isolate those aspects that matter for contagion dynamics (hereafter termed spreading dynamics when referring to specific dynamical models).  
We also summarize the topological features of higher-order structures utilized in the subsequent sections, and introduce the state spaces, dynamical observables underlying dynamics on higher-order networks.
The main notation used in this review is summarized in Table~\ref{tab:notation}.

\subsection{Simplicial complexes as geometric carriers of group interactions}

To model contagion processes with group interactions, one must specify how structural information beyond pairwise links is encoded and exposed to the dynamics. A standard network, represented as a graph \(G=(V,E)\), implicitly assumes that interactions occur only through dyadic contacts. Such a representation is insufficient when transmission depends on the simultaneous presence or joint influence of multiple individuals.

A \emph{simplicial complex} \(\mathcal{K}\)~\cite{iacopini2019simplicial,salnikov2018simplicial,wu2023simplicial,torres2020simplicial} is a collection of subsets of the node set \(V=\{1,\dots,N\}\). Each subset \(\sigma \subseteq V\), referred to as a \emph{simplex}, encodes a concrete group interaction involving all nodes in \(\sigma\), see Fig.~\ref{figure_model}. A defining feature of simplicial complexes is the closure property: whenever a group interaction (i.e., simplex) \(\sigma\) is present, all its non-empty subgroups \(\tau \subseteq \sigma\) are also included in \(\mathcal{K}\), reflecting the assumption that higher-order interactions are supported by their lower-order constituents. A simplex with \(|\sigma|=q+1\) nodes is referred to as a \(q\)-simplex. In this way, \(\mathcal{K}^{(0)}, \mathcal{K}^{(1)}, \mathcal{K}^{(2)},\dots\) correspond to individuals, pairwise interactions, three-body interactions, and higher-order group interactions, respectively~\cite{bianconi2021higher,shi2022simplicial}. The 1-skeleton \(G^{(1)}=(V,\mathcal{K}^{(1)})\) recovers the underlying contact network.

This hierarchical structure makes explicit how each node participates in interactions of different orders, which is a key ingredient for higher-order contagion models. For a node \(i\), we denote by \(k_i^{(q)}\) the number of \(q\)-simplices incident to \(i\), i.e., the number of group interactions of size \(q+1\) in which node \(i\) takes part. The vector $\mathbf{k}_i = \big(k_i^{(1)}, k_i^{(2)}, \dots \big)$ thus characterizes the multi-scale interaction environment experienced by node \(i\). 
In spreading dynamics, these quantities determine the number of potential transmission channels of different group sizes that are structurally available to an individual.
At the system level, the simplex order distribution \(|\mathcal{K}^{(q)}|\) specifies which interaction orders exist in the population and therefore constrains which higher-order transmission mechanisms can operate~\cite{iacopini2019simplicial}.

The closure property has an important physical interpretation: a \(q\)-simplex represents a group interaction embedded in a fully connected local neighborhood. For example, a 2-simplex \(\{i,j,\ell\}\) in Fig.~\ref{figure_model} implies that all three pairwise contacts among \(i\), \(j\), and \(\ell\) are present. This assumption captures settings in which group interactions emerge within cohesive social units, such as friendship cliques, co-authorship teams, or committees, rather than among loosely connected individuals~\cite{benson2018simplicial,chen2024simplicial,mulas2022graphs}. In the context of contagion, such cohesion can facilitate reinforcement, threshold effects, or collective exposure, which are crucial to many higher-order contagion mechanisms.

Building on this structure, mesoscopic descriptors of \(\mathcal{K}\), such as higher-order connected components or clustering measures (e.g., the fraction of edges participating in triangles), quantify where and how densely such cohesive groups are organized. These indicators identify regions of the system where higher-order transmission pathways are structurally concentrated and therefore where non-pairwise contagion effects are most likely to play a dominant role.

\begin{table}[!t]
  \caption{Main symbols and notation used in this review}
  \label{tab:notation}
  \centering
  \small
  \renewcommand{\arraystretch}{1.2}
  \setlength{\tabcolsep}{3pt}
  \begin{tabular}{p{0.18\columnwidth} p{0.8\columnwidth}}
    \hline
    \textbf{Notation} & \textbf{Definition} \\
    \hline

    $V$, $N = |V|$ &
    Node set of the system and its size. \\

    $G = (V,E)$ &
    Pairwise network (graph) with edge set $E$. \\

    $\mathcal{H} = (V,\mathcal{E})$ &
    Hypergraph with node set $V$ and hyperedge set $\mathcal{E}$. \\

    $e \in \mathcal{E}$, $|e|$ &
    Hyperedge and its size (number of nodes in $e$). \\

    $\mathcal{K}$, $\sigma$ &
    Simplicial complex $\mathcal{K}$ and a simplex $\sigma$ in $\mathcal{K}$. \\

    $\mathcal{T}$, $t$ &
    Set of time stamps and time index for temporal higher-order interactions. \\

    $\beta$ &
    Baseline (pairwise) infection rate. \\

    $\beta_k$ &
    Higher-order infection rate associated with simplex $\sigma$ with order $k$. \\

    $\lambda_{\text{eff}}$ &
    Effective spreading rate or control parameter (e.g., $\lambda_{\text{eff}} = \beta/\mu$, $\mu$ denotes the recovery rate). \\

    $f(\cdot)$ &
    Nonlinear infection kernel on hyperedges/simplices (e.g., group-level response function). \\

    $\lambda$ &
    Basic infection rate in hypergraph-based infection kernels.\\

    $\nu$ &
    Nonlinearity exponent in hypergraph-based infection kernels. \\

    $\theta_i$ &
    Threshold of node $i$ in threshold-activated contagion. \\

    $\epsilon$ &
    Coupling parameter between competing or interacting contagion processes. \\

    $\rho(t)$ &
    Infection (or adoption) density at time $t$. \\

    \hline
  \end{tabular}
\end{table}

\subsection{Hypergraphs as flexible models of group interactions}

Hypergraphs offer a more flexible representation of group interactions by relaxing the closure requirement imposed by simplicial complexes~\cite{estrada2005complex,antelmi2023survey,klamt2009hypergraphs,vazquez2023complex}, see Fig.~\ref{figure2}. A \emph{hypergraph} is defined as a pair $H=(V,\mathcal{E})$, where $V=\{1,\dots,N\}$ denotes the node set and $\mathcal{E}=\{e_1,\dots,e_M\}$ is a family of non-empty subsets $e_\alpha \subseteq V$, referred to as hyperedges~\cite{yamano2024hypergraph,wang2018development}. Each hyperedge represents a group interaction that directly couples all participating nodes, without assuming that interactions among all subgroups necessarily exist. Consequently, contagion may occur through simultaneous exposure within a group, rather than being mediated by an underlying dense network of pairwise or hierarchical contacts.

Two basic structural quantities are the \emph{hyperedge size} $s_\alpha = |e_\alpha|$, which quantifies the number of individuals jointly involved in a group interaction, and the \emph{node degree} $d_i = \big| \{\alpha : i \in e_\alpha\} \big|$, which counts how many distinct group interactions node $i$ participates in. In spreading dynamics, these quantities determine, respectively, the typical size of collective transmission events and the number of independent group-level transmission channels accessible to a node~\cite{ouvrard2020hypergraphs}. When all hyperedges have the same size, i.e., $s_\alpha = r$, the hypergraph is said to be $r$-uniform and represents a pure $r$-body interaction structure~\cite{cooper2012spectra,rodl2004regularity,nikiforov2014analytic,hu2015laplacian}.  In such systems, each interaction event involves exactly $r$ entities, so that the interaction order is homogeneous and pure $r$-body effects can be examined without interference from lower- or higher-order interactions. Consequently, $r$-uniform hypergraphs are frequently employed to isolate and study the dynamical effects of interaction order in higher-order spreading models.
In many applications, such as mailing lists, group chats, or task forces, hyperedges represent groups whose structure is defined by membership, meaning that some pairs inside the group may not interact directly; this is precisely what hypergraphs allow and simplicial complexes forbid.

In spreading dynamics, a few mesoscopic features of hyperedges play a crucial role in shaping spreading outcomes. In the following, we introduce three such patterns: hyperedge overlap, higher-order components, and hypergraph cores.
\emph{Hyperedge overlap}, the intersection size $|e_\alpha\cap e_\gamma|$, controls how strongly groups are coupled: high typical overlap yields dense corridors for gradual spreading, whereas low overlap isolates groups and favors more abrupt, avalanche-like activation~\cite{lee2021hyperedges,malizia2025hyperedge,lamata2025hyperedge}. \emph{Higher-order components}, obtained by chaining overlapping hyperedges, reveal regions where groups are tightly interwoven~\cite{kim2024higher,alt2005hocs}, while \emph{hypergraph cores}~\cite{mancastroppa2023hyper,bianconi2024nature,mancastroppa2024structural}, defined by iteratively pruning nodes in too few hyperedges, identify structurally central backbones of group interactions. Later sections will repeatedly show that the same contagion rule can exhibit smooth or explosive transitions, localized or global outbreaks, simply depending on these mesoscopic features of $H$.

\begin{figure}[t]
  \centering
  \includegraphics[width=\linewidth]{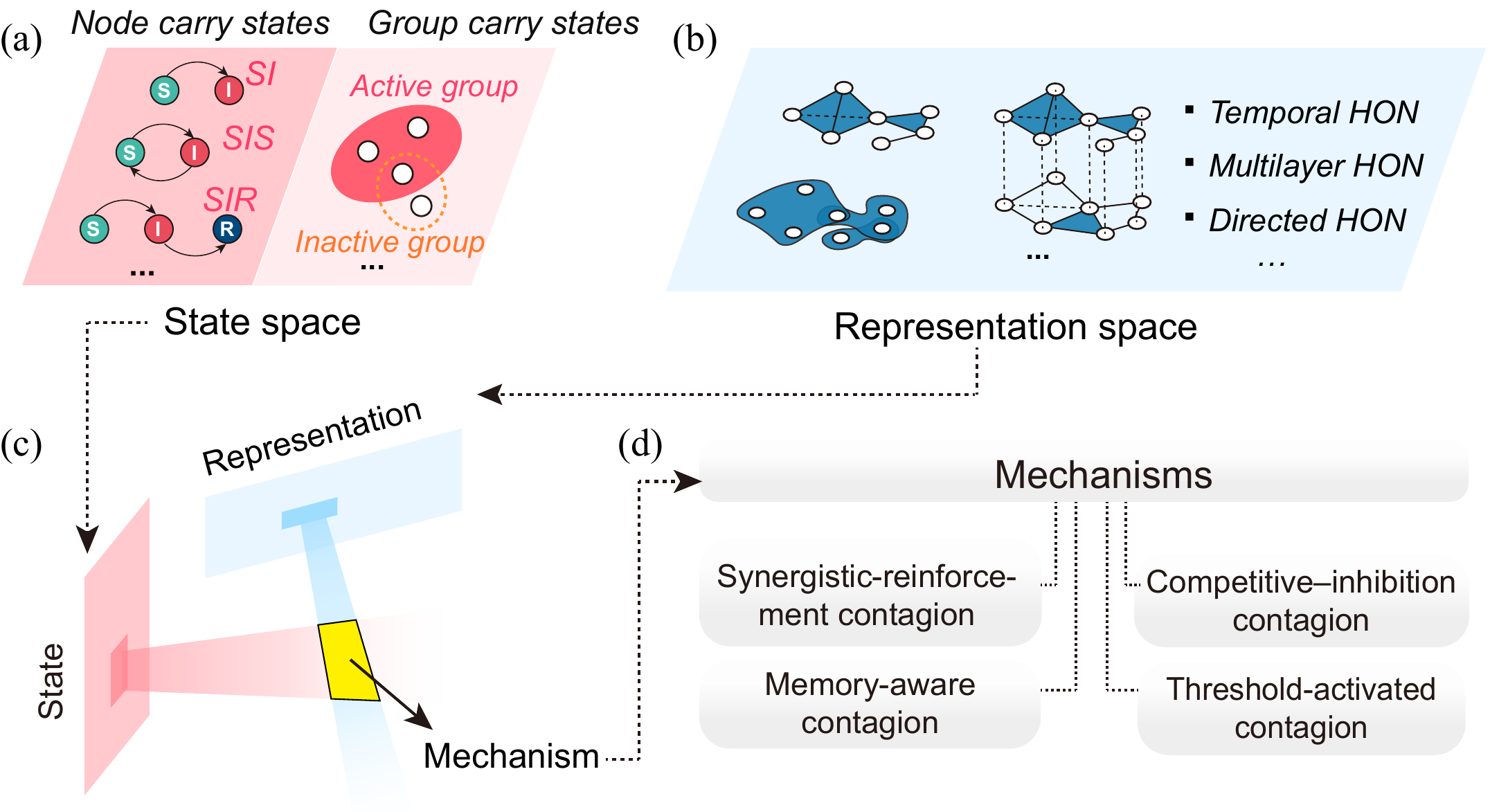}
  \caption{State spaces and interaction representations define the main categories of contagion mechanisms. (a) State spaces can be (i) node-based, such as Susceptible (S), Infected (I), and Recovered (R), or (ii) group-based, such as inactive and active. (b) Representations, including temporal, multilayer, and directed structures, are illustrated in Fig.~\ref{figure2} and described in Section~\ref{preliminaries}. (c) Mechanisms can be determined by two key choices: the state space and the representation of interactions. (d) In this work, we introduce the taxonomy and  focus on four main mechanism types.}
  \label{figure3}
\end{figure}

\subsection{Adding extra features to higher-order structures}
Real systems rarely exhibit higher-order structure in isolation. Group interactions are time-stamped, layered across domains, and may carry direction, sign or strength. These features do not replace simplicial complexes and hypergraphs but rather enrich them, yielding a family of \emph{temporal, multilayer and attributed higher-order networks} that more faithfully reflect empirical systems.

Temporal higher-order networks incorporate time by associating a timestamp or time window with each simplex or hyperedge~\cite{cencetti2021temporal,gallo2024higher,mancastroppa2024structural}. Instead of a static list of groups, one obtains a sequence of group interactions, each active only during a limited interval~\cite{myers2023topological,neuhauser2021consensus}. This time-resolved view fundamentally changes how contagion paths are defined: an infection can only follow a time-respecting chain of group interactions, and the ordering and temporal clustering of simplices or hyperedges becomes as important as their combinatorial pattern~\cite{petri2018simplicial,zhao2022effects}. Temporal simplicial complexes and temporal hypergraphs are thus natural substrates for contagion models with memory, burstiness and path dependence, which we revisit in Section~\ref{mechanisms} and Section~\ref{behaviors}.

Multilayer higher-order networks bring together several interaction contexts within a single framework.
Nodes may appear in multiple layers, each endowed with its own higher-order structure, such as simplicial complexes, hypergraphs, or their combinations~\cite{bai2025hyperbolic,wang2024identification}. 
For example, one layer may encode physical contacts through pairwise edges, another may encode awareness or opinion dynamics through simplicial complexes, and a third may encode groups sharing resources using hyperedges.
Inter-layer couplings connect copies of the same node across layers, allowing processes on different structural substrates to interact~\cite{krishnagopal2023topology,wang2025community,lotito2024multiplex}. This framework is particularly suited to contagion scenarios where a disease, an awareness process and a behavior spread on distinct but coupled higher-order structures, or where multiple technologies or norms coexist~\cite{fan2022epidemics,jalan2022multiple}.

Directional, signed and weighted higher-order interactions further refine the structural description. Directed simplices or hyperedges encode asymmetry in group interactions, such as broadcasting from a speaker to an audience, regulatory motifs in biochemical networks, or causally ordered events~\cite{pretolani2013finding, gallo1993directed, ausiello2017directed,masulli2016topology,gong2024higher}. Signed higher-order interactions distinguish between facilitating and inhibiting group influences, as in cooperative versus antagonistic group pressure, or positive versus negative feedback loops~\cite{chen2025dshgnn,chen2020neural}. Weights can quantify the intensity or frequency of group interactions~\cite{banerjee2020quantum,pan2025adaptive}. These attributes introduce structural avenues for heterogeneous influence, inhibition and competition at the group level, and will be essential when we discuss competitive contagion and signed cascades.

\subsection{Structural abstractions in higher-order contagion models}

When constructing higher-order contagion models, a key consideration is which structural features of the network should be explicitly incorporated, and how these features are translated into dynamical rules. Figure~\ref{figure3} illustrates that the mechanisms underlying higher-order contagion are also based on this rationale; namely, two modeling choices, (i) the state space and (ii) the representation of interactions, are particularly significant for mechanism construction.

The first concerns the \emph{state space}. In many models, only nodes carry states (such as susceptible, infected, recovered)~\cite{allen1994some, pastor2015epidemic,vespignani2012modelling}, while simplices or hyperedges are treated implicitly through their effect on node transition rates~\cite{iacopini2019simplicial,st2022influential}, as shown in Fig.~\ref{figure3}. In others, group entities themselves may have states (inactive versus active, uninfected versus infected hyperedges), and contagion proceeds through changes at both node and group levels~\cite{xie2023vital}. The latter approach is especially natural for threshold and activation-based models, where a hyperedge must first reach an “active” state before it can trigger changes in its member nodes.

The second concerns the \emph{representation of interactions}. One can choose an explicit higher-order representation, keeping simplices or hyperedges as primitive objects, or project them onto a pairwise network, for example by replacing each group with a clique of edges or with weighted co-occurrence links~\cite{wang2018critical}. Explicit representations preserve the distinction between different group sizes and configurations and make it straightforward to implement higher-order rules~\cite{antelmi2023survey}, such as infection that depends on the number of infected nodes in a hyperedge. Projections, by contrast, collapse group structure into effective pairwise statistics~\cite{yang2022semi,yan2024hypergraph, surana2022hypergraph}. They are simpler to handle analytically and computationally, but may suppress or distort phenomena that depend on the combinatorial identity of groups, including many of the discontinuous and multistable behaviors discussed later.

These modeling choices interact with the mechanisms that will be introduced in the next chapter. Synergistic reinforcement and threshold-activated contagion naturally exploit the group structure of simplices and hyperedges and are often formulated on explicit higher-order substrates. Competitive–inhibition contagion, including disease–disease and disease–awareness interactions, is most naturally described on multiplex higher-order networks, where different processes inhabit distinct but coupled structures. Memory-aware contagion finds its natural structural home on temporal simplicial complexes and temporal hypergraphs, where time-ordered group interactions and temporal correlations can be represented directly. In all cases, the expressive power of a contagion model depends not only on the mechanism it encodes but also on how faithfully the underlying higher-order structure is captured.

\subsection{Dynamical observables and basic notions}
Before turning to specific mechanisms and dynamical behaviors, it is useful to fix the basic observables and notions that will be used throughout the survey.

At the macroscopic level, the most common observable is the \emph{infection density} or prevalence, denoted $\rho(t)$, which measures the fraction of nodes that are infected or active at time $t$. In discrete-time models, one may also track the \emph{incidence}, that is, the number or fraction of newly infected nodes between times $t$ and $t+1$. For processes that terminate, such as SIR-type epidemics or irreversible adoption, the \emph{outbreak size} or final adoption size refers to the cumulative fraction of nodes that have ever been infected or adopted by the end of the process. These quantities are the natural order parameters for describing whether contagion dies out, remains localized, or reaches a large fraction of the system.

At the microscopic level, contagion rules typically involve one or more \emph{infection rates} or effective transmission parameters, often denoted $\beta$ or $\lambda$, which control how quickly infection or adoption spreads along allowed interaction channels. In higher-order models, such rates may depend on the order of the interaction (for example, distinct rates for edges, triangles, and higher simplices) or on the number of infected nodes inside a hyperedge. It is often convenient to summarize the combined effect of contact structure and infection rates in an \emph{effective reproduction number} or effective transmission parameter $\lambda_{\mathrm{eff}}$, which captures the average tendency of the process to grow or decay.

When analysing long-time behavior, one is typically interested in \emph{stationary states} or equilibria of the dynamics and their \emph{stability}. An equilibrium is said to be stable if small perturbations decay over time and unstable if they grow. A \emph{phase transition} occurs when a smooth change in a control parameter, such as an infection rate or a structural parameter, leads to a qualitative change in the number or nature of stable equilibria. In \emph{continuous} (second-order) transitions, the order parameter changes smoothly from zero and critical quantities such as susceptibility often diverge; in \emph{discontinuous} (first-order or hybrid) transitions, the order parameter exhibits a finite jump, frequently accompanied by bistability and hysteresis.

The present section provides the structural and conceptual vocabulary needed to interpret those results. In the sections that follow, these concepts will be instantiated on the higher-order substrates introduced here. Mechanisms such as synergistic reinforcement, competitive–inhibition, memory-aware contagion and threshold activation will be defined explicitly, and their interactions with the structural features of simplicial complexes, hypergraphs, temporal and multiplex higher-order networks will be traced through the resulting phase diagrams, multistability patterns, and propagation delays.

\section{Mechanisms of spreading dynamics}\label{mechanisms}
High-order interactions give rise to a diverse repertoire of structural patterns that fundamentally reshape the ways in which spreading processes unfold on higher-order networks.
Motivated by the limitations of the projections of higher-order networks, recent research has increasingly turned toward dynamical mechanisms that leverage the representational power of higher-order network formalisms. 
Some approaches rely primarily on the natural capacity of hyperedges, simplices, or motifs to encode group-level processes, whereas others enrich the modeling space through additional dimensions such as temporal correlations, multi-state transitions, or history-dependent rules.

\begin{figure}[t]
  \centering
  \includegraphics[width=.9\linewidth]{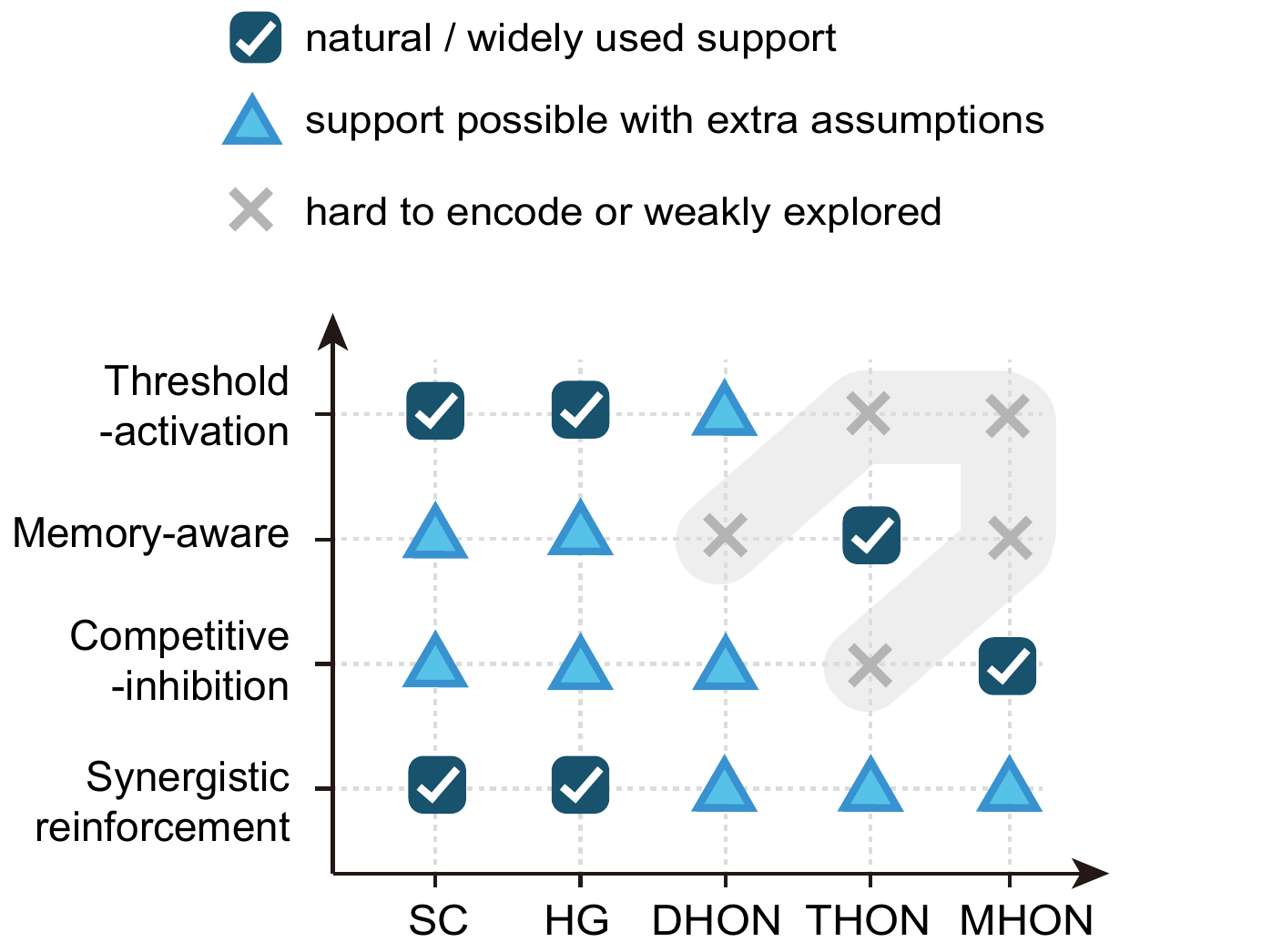}
  \caption{Structural carrying capacities of different higher-order network representations for distinct contagion mechanisms. The representations we focused on here including simplical complexes (SC), hypergraphs (HG), Directed higher-order networks (DHON), Temporal higher-order networks (THON) and Multilayer networks (MHON). Check-marked boxes indicate mechanisms that are naturally supported for the given structure and widely used in the literature, while triangles denote support that is possible only under additional modeling assumptions. Shaded or annotated regions highlight combinations with limited modeling evidence, marking current research gaps.}
  \label{figure4}
\end{figure}

\begin{figure*}[t]
  \centering
  \includegraphics[width=\linewidth]{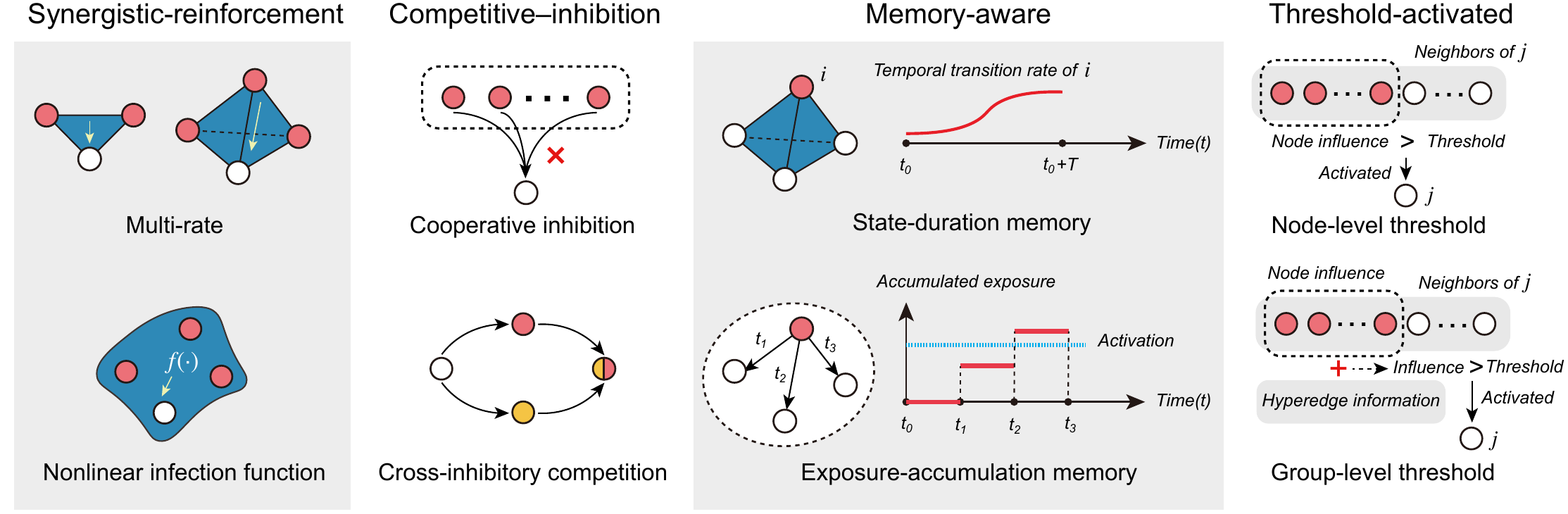}
  \caption{Illustration of the four classes of higher-order contagion mechanisms: (i) synergistic-reinforcement contagion, (ii) competitive–inhibition contagion, (iii) memory-aware contagion, and (iv) threshold-activated contagion. In the top panel for synergistic reinforcement, simplices encode order-dependent multi-rate higher-order channels, while in the bottom panel the function \(f(\cdot)\) denotes a nonlinear infection kernel defined on hyperedges in hypergraph-based contagion. In the competitive–inhibition panel, pink and yellow nodes indicate infected states under two distinct spreading dynamics. In the memory-aware and threshold-activated panels, vertices \(i\) and \(j\) denote the currently infected node and the to-be-activated node, respectively.
}
  \label{figure5}
\end{figure*}

This section synthesizes the major classes of contagion mechanisms developed for high-order networks, with an emphasis on how different higher-order structures differ in their capacity to support and express these dynamical mechanisms. We categorize the existing literature into four mechanism families: (i)\emph{synergistic-reinforcement contagion}, (ii) \emph{competitive-inhibition contagion}, (iii) \emph{memory-aware contagion}, and (iv) \emph{threshold-activated contagion}. For each class, we begin by outlining the core conceptual principles, then survey representative modeling approaches, and finally discuss how the structural features of different higher-order representations constrain or enhance the expressiveness of the corresponding dynamical mechanisms. Figure~\ref{figure4} summarizes the carrying capacity of various higher-order formalisms for these mechanisms, together with the available (or still limited) modeling evidence, while Fig.~\ref{figure5} provides a more detailed illustration of the mechanism families and their subcategories as organized in this review.

\subsection{Synergistic-reinforcement contagion}
Synergistic-reinforcement mechanisms represent one of the most prevalent classes of spreading dynamics on higher-order networks. These mechanisms are motivated by the intrinsically collective nature of group interactions, whereby the accumulation of infected individuals produces a nonlinear amplification of transmission, commonly characterized by power-law, exponential, or other nonlinear functional dependencies. Such nonlinear amplification captures the essence of group-driven reinforcement effects that cannot be represented in pairwise models.

\subsubsection{Classification of Synergistic-Reinforcement Contagion}

In higher-order settings, synergistic-reinforcement contagion can be implemented in multiple ways. Importantly, these implementations differ not only in modeling details, but in how the underlying interaction structure acts as a carrier of structural carrying capacity, i.e., in which structural layer collective reinforcement is encoded and activated. To make this distinction explicit, existing models can be broadly grouped into two complementary classes, as illustrated in Fig.~\ref{figure5}. The first class of models realizes synergy through \emph{order-dependent} and \emph{multi-rate higher-order channels} defined on simplices. In simplicial contagion models~\cite{iacopini2019simplicial,matamalas2020abrupt,wang2021simplicial}, a node receives transmissions both along edges and from higher-order simplices, with distinct infection rates $\beta_\sigma$ assigned to interactions of different orders (e.g. $\beta_\Delta$ and $\beta_\boxtimes$ for 2-simplex and 3-simplex in Fig.~\ref{figure5}, respectively). Effective group reinforcement appears when infection through a $q$-simplex is only allowed once a sufficiently large subset of its members is active. This class includes both \emph{collective contagion}, where full activation of the simplex is required, and \emph{individual contagion}, where the presence of any active neighbour in the simplex is sufficient, and it naturally captures \emph{critical mass}, like group effects on simplicial complexes.

The second class of models implements synergy via \emph{nonlinear infection-rate functions} on hyperedges. In these models, each hyperedge $e_\alpha$ contributes an infection pressure that depends nonlinearly on the number of active (infected) nodes $n_\alpha$ it contains, through a kernel $f(n_\alpha)$~\cite{de2020social,st2021universal,cencetti2023distinguishing,mancastroppa2023hyper}. Power-law, saturating or sigmoidal forms of $f(\cdot)$ have been used to encode super-linear reinforcement at small $n_\alpha$ and sub-linear effects at large $n_\alpha$, leading to multistability, hybrid transitions and hysteresis on general hypergraphs. Within this second class, inter-group transmission can follow different transport schemes on the hypergraph, such as reactive versus contact processes~\cite{suo2018information,xie2023efficient}, but these choices mainly specify how group-level infection attempts are scheduled in time rather than the nature of synergistic reinforcement itself.

These two families provide a unified view of synergistic-reinforcement contagion. Both encode that the effective infection rate experienced by a node is a nonlinear function of the number and configuration of active peers in its groups, but they do so using different structural carriers: simplices with interaction-order–dependent rates versus arbitrary hyperedges with explicit nonlinear kernels. In later sections, we show how these two implementations systematically underlie the emergence of multistability, discontinuous phase transitions and propagation inertia in higher-order contagion.

\subsubsection{Structural Carrying Capacity of Synergistic Reinforcement Mechanisms}
Synergistic-reinforcement mechanisms require higher-order structures that can encode either (i) cooperative infection events involving multiple nodes, or (ii) nonlinear infection-rate functions that explicitly capture such cooperative effects. These nonlinear reinforcement can be naturally supported by several families of higher-order architectures, including simplicial complexes and various forms of hypergraphs, each enabling distinct manifestations of multi-body interactions. In what follows, we examine how synergistic reinforcement can be embedded and extended within simplicial complexes, homogeneous and heterogeneous hypergraphs, and other specialized higher-order structures.

Simplicial complexes are composed of multi-node building blocks of different orders, making them a natural substrate for directly encoding cooperative infection. An influential simplicial contagion model~\cite{iacopini2019simplicial} introduces higher-order contagion by assigning to each simplex $\sigma$ ($|\sigma|=k+1$) a group-level infection rate $\beta_k$: if all nodes within a simplex adjacent to a susceptible node are infected, the node experiences both the baseline pairwise infection pressure $\beta$ and an additional simplex-induced reinforcement term. Subsequent works~\cite{landry2020effect} compared collective and individual contagion mechanisms and confirmed the mediating role of simplices in shaping higher-order spreading. Further extensions~\cite{chen2024simplicial} incorporated recovery processes, contact durations, and activity-driven temporal patterns to formulate SIR and SEIR dynamics on simplicial complexes for the study of COVID-19 control strategies. In these structures, nonlinear cooperative effects arise naturally from the presence of multi-node simplices.

Hypergraphs offer an alternative means of capturing synergistic reinforcement through nonlinear infection-rate functions. In uniform hypergraphs, contagion can be modeled by assigning identical $r$-size hyperedges specific transmission rates, as in~\cite{suo2018information}. More general formulations, such as those in~\cite{ferraz2021phase}, decompose arbitrary hypergraphs into layered $r$-uniform components, allowing spreading on any hypergraph to be expressed in a unified framework. For heterogeneous hypergraphs, many works define a susceptible node’s infection rate as a nonlinear function of the proportion of infected nodes in its incident hyperedges.
An SIS framework~\cite{higham2022mean} formalizes this idea by assigning to each node $i$ an instantaneous infection rate $\lambda_i = f(n_i)$ that encodes cooperative effects and satisfies $f(0)=0$, where $n_i$ is the number (or fraction) of infected nodes in the hyperedges incident to $i$.
Subsequent studies specify $f$ through particular nonlinear functional forms, most commonly power laws such as $f = \lambda (n_\alpha)^\nu$~\cite{st2024nonlinear,cencetti2023distinguishing}, or exponential-type reinforcement functions that saturate as the number of infected nodes in a hyperedge grows~\cite{mancastroppa2023hyper}. These formulations illustrate how hyperedge-level dependencies act as structural carriers of nonlinear synergy.

Moreover, some hypergraphs possess intrinsic structural regularities that inherently support synergistic contagion. The notion of higher-order components (HOCs), introduced in~\cite{kim2024higher}, defines $r$-order connectivity through shared nodes among hyperedges; nonlinear reinforcement is then expressed by assigning infection probabilities \(\beta_n\) within hyperedges containing \(n\) infected nodes. Related studies on nested hypergraphs~\cite{kim2023contagion,yang2025multilayer}, where smaller hyperedges are fully contained within larger ones, embed cooperative effects by specifying size- and load-dependent infection rates \(\beta(s,n)\), thus allowing hierarchical hyperedge organization to carry nonlinear contagion.

Synergistic reinforcement can also be embedded in directed, multilayer, and temporal hypergraphs, though additional constraints are required. Directed hypergraphs incorporate hyperedge directionality into the infection rules~\cite{li2024enhancing}. Multiplex hypergraphs allow inter-layer reinforcement through additive interactions or layer-dependent parameters~\cite{fan2022epidemics}. Temporal hypergraphs integrate time-resolved hyperedges, enabling cooperative effects to depend on temporal activation patterns~\cite{mancastroppa2025emerging,antelmi2020design,chowdhary2021simplicial,st2021universal}. Path-based structures, however, lack multi-node group interactions and thus cannot support synergistic reinforcement.

\subsection{Competitive–inhibition Contagion}
Competitive–inhibition contagion arises naturally in higher-order spreading processes when multi-node interactions generate exclusivity constraints, inhibitory influence, or resource conflicts. These situations produce diffusion dynamics in which transmission can be delayed, reduced, or blocked by the presence of competing pathways or suppressive group effects. Throughout this work, we use the term \emph{competitive–inhibition contagion} to collectively describe contagion processes in which higher-order interactions suppress or compete with spreading activity.

More concretely, competitive–inhibition contagion encompasses two broad situations: (i) group-induced suppression, where the local group context inhibits the transmission of a single contagion, and (ii) direct competition between multiple contagions, where several contagion processes are simultaneously coupled on the same higher-order structure. 
The former captures scenarios in which group-level factors discourage or delay transmission, even when individual-level conditions might favor it, whereas the latter reflects mutually exclusive or antagonistic contagions that vie for adoption, prevalence, or dominance within the same population.

\subsubsection{Classification of competitive–inhibition contagion}
Models of competitive–inhibition contagion can be broadly categorized according to the form of multi-node inhibitory influence. Here, the classification distinguishes models according to whether structural carrying capacity limits contagion through local saturation, resource competition, or global normalization of transmission rates.
The first class, \emph{cooperative inhibition}, comprises mechanisms in which groups of nodes jointly suppress the spread of a contagion during a higher-order interaction~\cite{st2022influential}. This class mirrors phenomena such as group hesitation, dissenting group opinions, or inhibitory social pressure that weaken or neutralize transmission. The second class, \textit{cross-inhibitory competition}, captures settings in which two or more contagion processes co-evolve on the same higher-order network and interact antagonistically~\cite{li2022competing,li2022two}. Here, a node may simultaneously hold distinct states with respect to different contagions.

\subsubsection{Structural carrying capacity of competitive–inhibition contagion}
The ability of higher-order structures to represent competitive–inhibition contagion depends on how effectively simplicial complexes, hypergraphs, and multilayer architectures can encode inhibitory or competitive interactions.
In simplicial complexes, cooperative inhibition is commonly implemented by reducing higher-order infection rates associated with simplices. Because simplices inherently encode multi-node interactions, assigning small or negative transmission rates \(\beta_k \le 0\) to higher-order simplices naturally models social inhibition, as demonstrated in~\cite{st2022influential}. Competitive inhibition can also be supported in simplicial complexes by allowing nodes to carry multiple contagion-specific states, thereby encoding cross-inhibitory competition between simultaneously spreading contagions. A competing irreversible contagion model~\cite{li2022two} considers two SIR-type processes, A and B, that simultaneously propagate on a simplicial complex. Nodes can occupy one of five states—susceptible (\(S\)), infected by A (\(I_A\)), infected by B (\(I_B\)), recovered from A (\(R_A\)), or recovered from B (\(R_B\))—so that the two contagions interact competitively. Similarly,~\cite{xue2022cooperative,lucas2023simplicially,lamata2024pathways} extend reversible SIS-like models by encoding three binary variables that represent infection by A, B, or both, and by governing competitive interactions through coupling factors \(\varepsilon_{AB}\) and \(\varepsilon_{BA}\). Values \(\varepsilon<1\) yield competition, \(\varepsilon>1\) cooperation, and \(\varepsilon=1\) independence.

Multilayer networks containing simplicial complexes further provide a natural way to carry cross-inhibitory
competition, since different layers may represent different spreading processes. Studies in~\cite{fan2022coupled,hu2024coupled} develop coupled disease–information diffusion models in which awareness propagates on a simplicial information layer while disease spreads on a physical-contact layer, allowing cross-layer competition or suppression. The same concept is extended in~\cite{wang2022interplay} by coupling a virtual-information layer with complex higher-order interactions to a temporal physical-contact layer, thereby capturing competition arising from time-varying group interactions.

In hypergraphs, cooperative inhibition can also be carried under specific nonlinear reinforcement settings. 
In classical hypergraph contagion models such as those proposed in~\cite{cencetti2023distinguishing,st2024nonlinear}, setting the nonlinear scaling exponent $\nu < 1$ produces sublinear infection reinforcement, which functions analogously to social inhibition in simplicial complexes.
cross-inhibitory
competition on hypergraphs has been less explored, with notable exceptions such as bi-virus dynamics on directed hypergraphs~\cite{cui2023general,gracy2024competitive}. 
In these models, two interacting contagions propagate through directed hyperedges, with the directionality of hyperedges and epidemic parameters jointly shaping which contagion dominates or whether coexistence emerges.
Despite the rich potential of hyperedges to encode multi-node exclusivity or resource constraints, multilayer hypergraph models explicitly incorporating competitive–inhibition contagion remain largely undeveloped.
Overall, competitive–inhibition contagion can be embedded in higher-order structures by assigning inhibitory transmission functions, allowing multi-process node states, or incorporating cross-layer coupling. While simplicial complexes and their multilayer variants have supported substantial progress, competitive inhibition in general hypergraphs remains an open avenue for future research.

\subsection{Memory-Aware Contagion}
Memory-aware contagion mechanisms represent a fundamental departure from classical Markovian spreading processes, where future evolution is fully determined by the current state of the system. In memory-aware dynamics, the temporal history of past interactions, the ordering of exposures, or the cumulative effect of repeated signals explicitly influences state transitions, thereby introducing non-Markovian features into the spreading process. Such mechanisms naturally arise in systems exhibiting behavioral inertia, delayed activation, habituation effects, or exposure thresholds that depend not solely on instantaneous configurations but on the timing and sequence of prior interactions. Notably, the incorporation of memory breaks the equivalence between static and temporal representations of the underlying interaction structure: propagation unfolds along temporal paths rather than purely structural ones, often producing early-time bottlenecks, delayed outbreak phases, or persistent propagation inertia. These phenomena underscore the central role of memory in shaping macroscopic spreading outcomes on higher-order and temporal networks.

\subsubsection{Classification of Memory-Aware Contagion}
Memory-aware spreading on higher-order networks can be divided into two main categories according to how temporal history enters the transmission process.
This distinction thereby clarifies whether structural carrying capacity is enforced instantaneously by current group configurations or accumulated over time through repeated exposures.
The first category, \textit{state-duration memory}, refers to models where a node’s transition probabilities depend on its past trajectory~\cite{lin2024higher}. Examples include the time spent in an infected state, previous activations, or the interval since the last exposure. Here, memory modifies transition rates through time-dependent state evolution, allowing the process to capture delayed activation, prolonged infectiousness, or gradual recovery. The second category, \textit{exposure-accumulation memory}, links transmission to the history of past contacts rather than the duration of a node’s state~\cite{gallo2024higher}. Adoption or infection occurs only after repeated, ordered, or temporally clustered exposures. In this case, memory is encoded through reinforcement rules that accumulate influence over time, reflecting situations where single exposures are insufficient and coordinated signals are required to trigger activation.

\subsubsection{Structural carrying capacity of memory-aware contagion}
The extent to which a network can sustain memory-dependent contagion is closely tied to its intrinsic combinatorial and temporal organization. In simplicial complexes, memory is often implemented through non-Markovian dynamics acting on higher-order simplices~\cite{lin2025impacts,lin2024higher}, where transmission or recovery probabilities depend on a node’s prior exposures within multi-node interaction groups. This enables simplicial complexes to encode propagation history at the level of higher-order interactions, naturally supporting cumulative exposure effects and sequential state dependence. In temporal hypergraphs~\cite{gallo2024higher}, memory can be incorporated through group-based temporal windows, where the activation or reactivation of hyperedges depends on recent group activity patterns, thereby capturing collective forms of memory across repeated higher-order contacts. These two structures thus offer complementary ways of embedding memory, i.e., simplicial complexes emphasize node-level temporal dependence within higher-order interactions, while hypergraphs emphasize aggregated group-level temporal memory.

Beyond these settings, multiplex networks with higher-order links provide additional opportunities to embed memory through dynamically evolving cross-layer interactions, where a node’s activity or exposure in one layer influences its susceptibility or behavior in another. However, the study of memory-aware contagion in general higher-order, multilayer, or hybrid structures remains limited. The degree to which these richer architectures can sustain complex path-dependent dynamics, i.e., a hallmark of real-world behavioral, biological, and social contagion, remains an important open challenge in higher-order network science.

\subsection{Threshold-activated contagion}
Threshold-activated contagion describes spreading processes in which activation requires a node, or an entire group, to receive influence exceeding a predefined threshold. Unlike classical pairwise models, threshold mechanisms formalize group-level effects through explicit activation rules, enabling quantitative analysis of cascades, critical transitions, and tipping behavior. These mechanisms naturally generalize synergistic contagion by imposing sharp, nonlinear activation conditions that depend on collective input rather than single contacts. Threshold models have been applied extensively to simplicial complexes and hypergraphs, revealing phenomena such as abrupt phase transitions, cascade criticality, and multi-stable regimes. By linking micro-level adoption rules to mesoscopic group interactions, threshold mechanisms provide a flexible framework for studying how structural organization and higher-order interactions jointly shape spreading outcomes.

\subsubsection{Classification of threshold-activated contagion}
Threshold models on higher-order networks can be classified along two main dimensions, both of which determine where and how structural carrying capacity shapes spreading. The first concerns where the threshold operates. \textit{Node-level threshold models} activate a node once the influence from its neighbors surpasses a predefined threshold, capturing individual-level decision rules~\cite{xie2023universal}. \textit{Group-level threshold models}, by contrast, apply thresholds to hyperedges or simplices, so that activation depends on the joint state of a multi-node interaction. This distinction reflects whether the critical condition is individual or collective~\cite{kilic2022simplicial,xie2023vital}. The second axis concerns the form of the threshold. \emph{Single-threshold models}, which assign each unit a fixed (possibly heterogeneous) threshold, dominate current research. \emph{Multi-threshold models}, in which thresholds evolve over time, depend on memory, or vary with exposure history, remain largely unexplored but hold potential for capturing richer dynamics in higher-order systems.

Concretely, in a standard node-level rule on graphs~\cite{granovetter1978threshold,watts2002simple}, node $i$ becomes active when the fraction of its active neighbors exceeds a threshold $\theta_i$, i.e., $ \frac{1}{D_i}\sum_{j \in \mathcal{N}(i)} \mathbb{I}\{X_j = 1\} \;\ge\; \theta_i$,
where $D_i$ is the degree of $i$, $\mathcal{N}(i)$ its neighborhood, and $\theta_i \in [0,1]$ may be homogeneous or heterogeneous across nodes. On simplicial complexes, thresholds can depend on simplex order, so that activation compares the fraction of active neighbors within $k$-simplices against an order-specific threshold $\theta_k$, as in the simplicial threshold model (STM)~\cite{kilic2022simplicial}. On hypergraphs, threshold rules are typically expressed at the hyperedge level, with node $i$ activating once the number or fraction of active members in a hyperedge $e$ containing $i$ exceeds a hyperedge-specific threshold $\theta_e$~\cite{xie2023universal}. These canonical forms illustrate how threshold conditions can be attached to nodes, simplex orders, or individual hyperedges.

\subsubsection{Structural carrying capacity of threshold-activated contagion}
A network’s ability to sustain threshold-based contagion depends strongly on its higher-order structure. Simplicial complexes naturally support threshold dynamics because order-dependent thresholds $\theta_k$ can be defined on simplices of different orders. The STM model on simplicial complex~\cite{kilic2022simplicial} demonstrates how higher-order geometry can govern cascade timing, delay propagation, and induce inertia, showing that simplex structure directly shapes threshold-driven spreading. Hypergraphs offer a more flexible substrate: hyperedge-level thresholds $\theta_e$ enable heterogeneous group activation and support computational tasks such as influence maximization~\cite{xie2023universal}. Studies on undirected hypergraphs~\cite{xu2022dynamics,zhang2024influence,ferraz2023multistability} show that these structures can be tuned to express rich families of group-level thresholds while retaining analytical tractability. Directed hypergraphs further allow asymmetric activation rules, for example by assigning different thresholds to incoming and outgoing hyperedges, leading to dynamics that are sensitive to hyperedge directionality~\cite{xu2025estimation}. 

More complex structures introduce additional challenges. Temporal hypergraphs encode time-varying group interactions, but the shifting interaction context makes fixed threshold definitions harder to apply and often calls for time-dependent or history-dependent thresholds. Multiplex networks with higher-order links, in principle, allow layer-specific thresholds $\theta_i^{[\ell]}$ and inter-layer coupling, offering considerable expressive power, yet systematic exploration of threshold models in these settings remains limited.

\begin{figure*}[t]
  \centering
  \includegraphics[width=\linewidth]{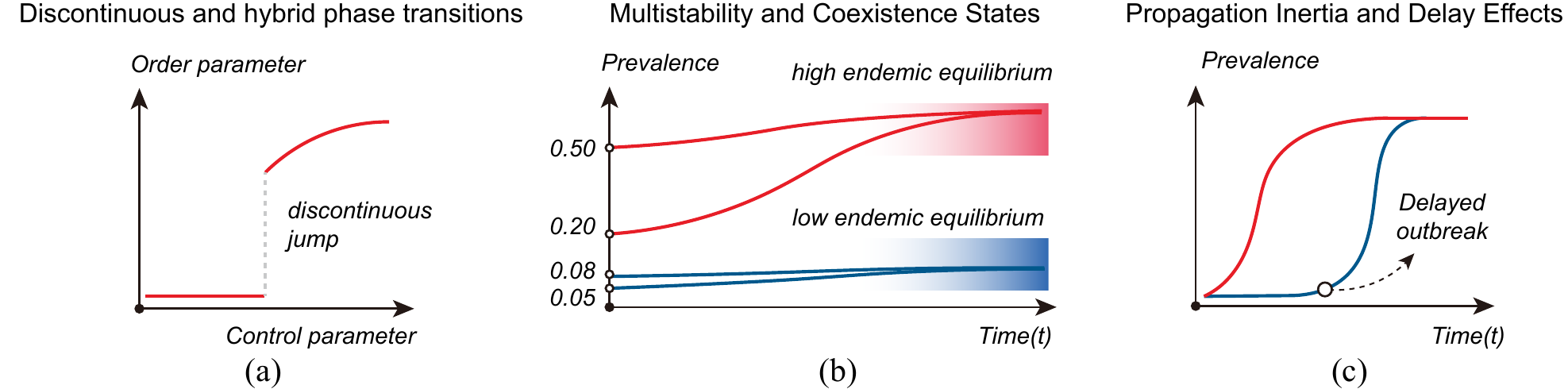}
  \caption{Illustrative dynamical regimes in higher-order contagion.
  (a) Discontinuous and hybrid phase transitions: as a control parameter (e.g. effective spreading rate) is varied, the order parameter displays a finite jump between low- and high-prevalence branches, with a bistable region and hysteresis between forward and backward parameter sweeps.
(b) Multistability and coexistence: for fixed parameters, the steady-state prevalence depends on the initial infected fraction, with both low and high endemic equilibria coexisting over a range of initial conditions.
(c) Propagation inertia and delayed outbreaks: time series of prevalence showing a long latent plateau followed by a sudden outbreak, contrasted with a rapid outbreak, highlighting how higher-order structures and mechanisms reshape transient spreading dynamics.}
  \label{figure6}
\end{figure*}

\begin{table*}[!t]
  \caption{Key dynamical concepts related to the three emergent behaviors}
  \label{tab:dyn_glossary}
  \centering
  \small
  \renewcommand{\arraystretch}{1.2}
  \setlength{\tabcolsep}{3pt}
  \begin{tabular}{p{0.28\textwidth} p{0.62\textwidth}}
    \hline
    \textbf{Dynamical concept} & \textbf{Interpretation in contagion dynamics} \\
    \hline

    Macroscopic state &
    System-wide configuration of the contagion, typically summarised by an aggregate quantity such as infection density $\rho(t)$. \\

    Steady state / equilibrium &
    Macroscopic state that does not change in time under the dynamics (up to small fluctuations). \\

    Endemic state &
    Steady state with a persistent, non-zero level of infection in the population. \\

    Basin of attraction &
    Set of initial conditions that converge to the same stable macroscopic state. \\

    Order parameter &
    Macroscopic observable used to distinguish regimes, such as infection density or final outbreak size. \\

    Hybrid phase transition &
    Transition combining an abrupt jump with continuous critical features (e.g., square-root scaling near the threshold). \\

    Critical point / threshold &
    Parameter value at which a phase transition occurs. \\

    Hysteresis &
    Dependence of the macroscopic state on the direction of parameter change (e.g., different thresholds when sweeping up vs down). \\

    Tipping point &
    Condition beyond which small perturbations can trigger a sudden, system-wide change. \\

    Plateau / quasi-stationary plateau &
    Extended period with nearly constant intermediate activity before relaxation to the final state. \\

    \hline
  \end{tabular}
\end{table*}

\section{Emergent Dynamical Behaviors}\label{behaviors}
In this section, we examine how specific combinations of higher-order structures and contagion mechanisms give rise to emergent dynamical behaviors. Specifically, we focus on discontinuous and hybrid phase transitions, multistability and coexistence states, and propagation inertia and delay effects, which are now widely reported in dynamics on higher-order networks. The main dynamical concepts used in this section are summarized in Table~\ref{tab:dyn_glossary}. Table~\ref{tab:structure_mechanism_mini} then provides a compact mapping between mechanisms, structural representations, and these three behavioral regimes, and the subsections below examine each regime in detail.

\subsection{Discontinuous and hybrid phase transitions}
Discontinuous phase transitions are recurrent features of contagion on higher-order networks. Unlike continuous transitions where the order parameter grows smoothly from zero (as is common in graph-based contagion models), here the steady-state infected or adopted fraction changes by a finite jump as a control parameter crosses a critical value, often accompanied by bistability and hysteresis, see Fig.~\ref{figure6}. On simplicial complexes and hypergraphs, such abrupt changes have now been observed in several distinct contagion families, and they can be traced to how group-level interactions, competition, memory effects and thresholds are embedded in higher-order structures.

The clearest body of evidence comes from synergistic reinforcement. On simplicial complexes,~\cite{iacopini2019simplicial} shows that when infection through 2-simplices is strong enough, the transition out of the absorbing state is no longer continuous: the stationary adoption density jumps from almost zero to a finite plateau. Building on the same setting, an SIS epidemic on simplicial complexes undergoes an abrupt change of regime at a tricritical point~\cite{matamalas2020abrupt}: for weak simplicial effects the endemic state appears via a continuous transition, while beyond that point the endemic fraction emerges through a hybrid transition with a finite jump and diverging susceptibility. ~\cite{wang2021simplicial} extends this analysis to a simplicial SIRS model with nonlinear incidence and find that the discontinuous transition and associated bistable branch persist, and that additional oscillatory outbreaks develop beyond the first-order transition. Moving from simplicial complexes to uniform hypergraphs,  a simplicial SIS process on scale-free uniform hypergraphs again exhibits a hybrid contagion transition when higher-order interactions are sufficiently strong~\cite{jhun2019simplicial}, i.e., the order parameter changes discontinuously at a finite threshold while critical fluctuations still diverge, combining first- and second-order features.

Synergistic reinforcement contagion on more general hypergraphs reinforces this picture and clarifies the structural ingredients behind it. A social contagion model on hypergraphs proposed in~\cite{de2020social} shows that increasing the nonlinearity of the group response drives the system from a continuous to a discontinuous transition: in the strongly nonlinear regime, the final adoption fraction exhibits a finite jump and the parameter space contains a multistable region in which low- and high-adoption states coexist. In~\cite{landry2020effect}, SIS-type dynamics on hypergraphs are analyzed, and it is shown that adding higher-order terms to an otherwise standard model can transform a simple transcritical bifurcation into a discontinuous transition with bistability, even when the underlying hypergraph is structurally simple. In a more general framework,~\cite{st2021universal} derive a nonlinear infection kernel for SIS dynamics on hypergraphs by combining heterogeneous exposure and a minimal infective dose, and show that under this kernel endemic prevalence displays discontinuous transitions, super-exponential growth near onset and pronounced hysteresis.~\cite{malizia2025hyperedge} then connect these dynamical features directly to mesoscopic structure by introducing a measure of hyperedge overlap and showing that explosive, abrupt transitions and bistability occur only when intra-order overlap is sufficiently low; as overlap increases past a critical value, the same contagion rules give rise to a smooth, continuous transition.

Discontinuous behavior also appears in competitive–inhibition contagion, although its manifestation is different: here the discontinuity often concerns which contagion dominates, rather than the existence of an endemic state per se. On simplicial complexes,~\cite{li2022competing} consider two competing SIS-like processes and show that the phase diagram splits into regions where one epidemic dominates, regions where the other dominates, and regions where both dominance outcomes are possible depending on initial conditions. They find that, as the infection strength associated with 2-simplices is varied, the system crosses sharp phase boundaries and the long-time outbreak size of each process changes non-smoothly, giving rise to discontinuous transitions between competing endemic states with hysteresis. In a related SIR-type framework on simplicial complexes~\cite{zhao2024susceptible}, the same group shows that strengthening higher-order transmission reshapes the phase portrait from one with a single dominant epidemic to regimes where either epidemic, or neither, can prevail, again with sharp changes in final epidemic size across the corresponding boundaries. On hypergraphs, ~\cite{gracy2024competitive} analyze a competitive bivirus SIS model with both pairwise and higher-order interactions and show that there are parameter regions with three locally stable equilibria: the disease-free state and two distinct single-virus endemic states. In addition to establishing the existence of three equilibria, this work further demonstrates that all three can be locally exponentially stable simultaneously.

For memory-aware contagion, explicit results on the order of the phase transition in higher-order settings are fewer, but they point to an important modulatory role for memory. ~\cite{lin2024higher} study a non-Markovian social contagion model on simplicial complexes by introducing non-exponential waiting times into a higher-order adoption process and mapping it to a Markovian system with auxiliary nodes. They show that increasing the average 2-simplicial degree can change the transition from continuous to discontinuous: when higher-order connectivity is weak, the steady-state infection density grows smoothly with the control parameter, whereas for stronger simplicial connectivity the same system exhibits a finite jump and hysteresis when parameters are swept forward and backward. In this case, memory does not create discontinuity in isolation, but it shifts the balance between activation and recovery so that the same higher-order system can sit either in a second-order or a first-order regime depending on how strongly non-Markovian effects and simplicial interactions are expressed. Non-Markovian epidemic models on temporal networks without explicit higher-order structure show that heavy-tailed inter-event time distributions can sharpen or broaden critical regions. Numerical studies combining such temporal memory with simplicial contagion suggest that explosive transitions identified in Markovian models may be enhanced or suppressed. However, a systematic classification of memory-induced first-order transitions on temporal hypergraphs remains open.

\begin{table*}[!t]
  \caption{Qualitative tendencies of higher-order contagion mechanisms on different structures in terms of three emergent behaviors. Shaded rows highlight distinct categories of higher-order networks.}
  \label{tab:structure_mechanism_mini}
  \centering
  \footnotesize
  \renewcommand{\arraystretch}{3.5}  
  \setlength{\tabcolsep}{4pt}

  \rowcolors{2}{gray!10}{white}  

  \begin{tabular}{p{0.18\textwidth} p{0.23\textwidth} p{0.19\textwidth} p{0.14\textwidth} p{0.19\textwidth}}
    \hline
    \rowcolor{gray!15}
    \diagbox[height=5.2em]{Structure}{Mechanism} &
    \textbf{Synergistic} &
    \textbf{Competitive--inhibition} &
    \textbf{Memory-aware} &
    \textbf{Threshold-activated} \\
    \hline

    \textbf{Simplicial complexes} &
    \shortstack[l]{Multistability / coexistence\\Discontinuous or hybrid transitions} &
    \shortstack[l]{Coexisting endemic states\\Discontinuous switching} &
    \shortstack[l]{Propagation inertia\\Delayed outbreaks} &
    \shortstack[l]{Discontinuous / hybrid cascades\\Abrupt activation} \\
    
    \textbf{Hypergraphs} &
    \shortstack[l]{Multistability / intermittency\\Hybrid transitions} &
    \shortstack[l]{Coexistence and dominance\\Switching (tristability)} &
    \shortstack[l]{Slow relaxation\\(potential inertia)} &
    \shortstack[l]{Hybrid percolation-like\\Discontinuous transitions} \\

    \shortstack[l]{\textbf{Temporal higher-order}\\\textbf{networks}} &
    \shortstack[l]{Delayed onset of\\Reinforcement (inertia)} &
    \shortstack[l]{Slow dominance selection\\between processes} &
    \shortstack[l]{Long plateaus\\Delayed outbreaks} &
    \shortstack[l]{Timing-sensitive activation\\and delayed cascades} \\

    \shortstack[l]{\textbf{Multiplex higher-order}\\\textbf{networks}} &
    \shortstack[l]{Cross-layer synergy\\Supporting multistability} &
    \shortstack[l]{Coexistence and layer-wise\\Discontinuous switching} &
    \shortstack[l]{Cross-layer memory\\Induced inertia} &
    \shortstack[l]{Multilayer tipping points\\Discontinuous transitions} \\
    \hline

  \end{tabular}
\end{table*}

Threshold-activated contagion on higher-order structures links discontinuous transitions to group-level activation criteria. On simplicial community networks,~\cite{li2024social} propose a threshold model and show that the final adoption size as a function of connection density and threshold parameters can undergo either smooth changes or abrupt jumps; in particular, they identify parameter regimes where cascade size increases continuously and regimes where small parameter shifts trigger large jumps in the fraction of active nodes. On signed hypergraphs~\cite{wang5287803information}, recent work shows that the cascade size as a function of average hyperdegree undergoes a continuous transition for negative values of a coupling parameter, but a discontinuous, first-order jump for non-negative values, with a clear separation between small-cascade and large-cascade phases. Bootstrap- and $k$-core-type percolation processes on hypergraphs~\cite{wang2022generalized}, where activation depends on satisfying local group thresholds, display analogous hybrid transitions: the size of the surviving core or active set drops discontinuously at a finite threshold and exhibits critical scaling around the transition point. These structural percolation results mirror what threshold-type contagion rules can produce dynamically when implemented on simplices and hyperedges.

\subsection{Multistability and Coexistence States}
Multistability, where a system admits multiple stable long-term outcomes under identical structural and dynamical settings, is a hallmark of higher-order contagion. Unlike classical pairwise spreading, which typically converges to a single dominant equilibrium, higher-order interactions generate nonlinear feedback mechanisms that can sustain distinct stable adoption levels and coexistence regimes (as shown in Fig.~\ref{figure6}), with the realized outcome determined by initial conditions and perturbations.

In synergistic contagion, multistability arises primarily from nonlinear reinforcement at the group level. Higher-order infection terms in simplicial complexes and hypergraphs, such as the simplex-based contagion introduced in~\cite{iacopini2019simplicial} and the nonlinear hyperedge scaling considered in~\cite{st2024nonlinear,cencetti2023distinguishing}, induce strong positive feedback that stabilizes both low-activity and high-activity equilibria. Simplicial complexes further amplify this effect, as higher-order simplices generate parameter regions in which both a healthy (inactive) equilibrium and an endemic (high-activity) equilibrium are simultaneously stable, so that the long-term outcome is determined by initial conditions, as shown in~\cite{iacopini2019simplicial} and in the subsequent analyzes in~\cite{petri2018simplicial}. Analogous multistable behaviors have been identified in hypergraph-based social contagion models, where the adoption probability depends nonlinearly on the fraction of active nodes within each hyperedge. In particular, the study~\cite{de2020social} demonstrated that social contagion on hypergraphs can display coexisting low-activity and high-activity equilibria, highlighting that hyperedge-level nonlinearities provide a generic route to multistability in higher-order spreading processes.

Competitive–inhibition contagion produces multistability through antagonistic interactions. Competing contagion models on higher-order structures, such as the epidemic-coupling SIR processes on simplicial complexes proposed by~\cite{li2022two} or the bi-virus dynamics on directed hypergraphs examined by~\cite{cui2023general,gracy2024competitive}, exhibit multiple stable outcomes, including epidemic-free state, dominance of either contagion, and, in some parameter regimes, coexistence equilibria in which both contagions persist. Extensions of reversible competing SIS-like processes on hypergraphs and networks, as studied in~\cite{xue2022cooperative,lucas2023simplicially,lamata2024pathways} further show how higher-order connectivity enlarges the state space and stabilizes asymmetric coexistence outcomes.

Threshold-activated contagion can also display multistability, although it has been analyzed less systematically than in synergistic reinforcement models. Even in classical pairwise settings, the Watts threshold model~\cite{watts2002simple} admits parameter regimes in which the same microscopic threshold rules can lead either to negligible activation or to large-scale cascades. Higher-order threshold mechanisms sharpen this effect. A simplicial threshold model for social contagion on higher-order community networks, proposed in~\cite{li2024social}, combines vertex-level adoption thresholds with higher-order community structure and reveals a clear hysteresis loop in the final adoption level as a function of the control parameters. This hysteresis implies the coexistence of low-adoption and high-adoption equilibria for the same parameter values, with the realized outcome selected by the initial conditions. These results indicate that threshold-activated contagion on higher-order structures can naturally support multiple competing macroscopic steady states, even though a full fixed-point classification of multistable regimes for general higher-order threshold models remains an open direction.

Memory-aware contagion provides another route to multistability by making spreading explicitly history dependent. On pairwise networks, social contagion models with cumulative-exposure memory have been shown to exhibit discontinuous transitions and hysteresis loops, so that low- and high-adoption equilibria coexist for the same parameters and the final adoption level is selected by initial seeding~\cite{wang2015dynamics,su2017emergence}. Extending these ideas to higher-order structures, a higher-order non-Markovian social contagion model on simplicial complexes proposed in~\cite{lin2024higher} incorporates memory kernels at the simplex level and shows that non-Markovian recovery can substantially modify steady states and system resilience, suggesting that the combination of higher-order group interactions and memory provides a natural mechanism for rich, potentially multistable spreading dynamics on higher-order networks.

\begin{table*}[!t]
\centering
\renewcommand{\arraystretch}{1.5} 
\setlength{\tabcolsep}{7pt}       
\setlength{\extrarowheight}{4pt}  

\caption{Mechanism--structure interactions and emergent behaviors in higher-order contagion. Only dominant and contributing mechanisms are indicated, where $\checkmark\checkmark$ denotes primary drivers and $\checkmark$ denotes secondary contributions. The mechanisms are synergistic-reinforcement (SR) contagion,  competitive–inhibition (CI) contagion, memory-aware (MA) contagion, and threshold-activated (TA) contagion.}
\label{tab:mech_struct_behaviors}

\begin{tabular}{m{0.17\textwidth} c c c c m{0.27\textwidth} m{0.28\textwidth}}
\hline
\rowcolor{gray!35}
\textbf{Emergent behaviors} & \multicolumn{4}{c}{\textbf{Dominant mechanisms}} & \textbf{Structural amplifiers} & \textbf{Typical signatures} \\
\rowcolor{gray!10}
& \textbf{SR} & \textbf{CI} & \textbf{MA} & \textbf{TA} & & \\
\hline

\rowcolor{blue!4}
\addlinespace[6pt]
\shortstack[l]{Multistability \\ \& coexistence} &
$\checkmark\checkmark$ & $\checkmark$ & $\checkmark$ &  & 
\shortstack[l]{Simplicial complexes (strong HO terms);\\
Hypergraphs (nonlinear kernels, low overlap);\\
Higher-order cores/components} &
\shortstack[l]{Multiple stable states;\\
Coexisting endemic levels;\\
Initial-condition sensitivity;\\
Long-lived coexistence} \\
\addlinespace[6pt]
\hline
\rowcolor{green!4}
\addlinespace[6pt]
\shortstack[l]{Discontinuous / hybrid \\ phase transitions} &
$\checkmark\checkmark$ & $\checkmark$ &  & $\checkmark\checkmark$ &
\shortstack[l]{Simplicial-dominated interactions;\\
Hypergraphs (convex/nonlinear kernels);\\
Hyper $k$-cores, bootstrap structures;\\
Signed/directed motifs} &
\shortstack[l]{Abrupt prevalence jumps;\\
Hybrid transitions;\\
Hysteresis \& bistability;\\
Tricritical regions} \\
\addlinespace[6pt]
\hline
\rowcolor{orange!4}
\addlinespace[6pt]
\shortstack[l]{Propagation inertia \\ \& delayed outbreaks} &
$\checkmark$ & $\checkmark$ & $\checkmark\checkmark$ & $\checkmark$ &
\shortstack[l]{Strong higher-order terms;\\
Temporal correlations;\\
Multiplex HO coupling;\\
Core--periphery structure} &
\shortstack[l]{Long plateaus;\\
Delayed activation;\\
Slow attractor selection;\\
Suppressed or delayed outbreaks} \\
\addlinespace[6pt]
\hline
\end{tabular}
\end{table*}

\subsection{Propagation Inertia and Delay Effects}
Beyond the impact on steady states and phase transitions, higher-order contagion models often exhibit pronounced propagation inertia: slow initial growth, long-lived low-activity plateaus, delayed outbreaks, and sluggish relaxation back to healthy states, see Fig.~\ref{figure6}. These effects arise from the same structural and mechanistic ingredients that drive multistability and hybrid transitions, including nonlinear group reinforcement, competition between processes, memory, and thresholds, but they manifest at the level of transient trajectories rather than only in asymptotic phases. Recent work has begun to chart when such delays occur, and how they depend on the interplay between higher-order structure and the four mechanism classes discussed in Section~\ref{preliminaries}.

In synergistic contagion, inertia emerges because higher-order reinforcement requires a sufficient number of infected neighbors within a simplex or hyperedge before amplification occurs. In the simplicial contagion model~\cite{iacopini2019simplicial}, low initial prevalence fails to activate higher-order terms, causing a prolonged “slow-spread” phase before nonlinear reinforcement kicks in. 
On hypergraphs, ~\cite{ferraz2024contagion} and~\cite{st2021universal} show that nonlinear social contagion with group reinforcement can enter regimes of intermittency, in which the global adoption level spends long time intervals close to either low- or high-activity plateaus and only occasionally makes sudden excursions between them.
A complementary perspective is given by~\cite{malizia2025disentangling}, who disentangle the roles of hyperedge size heterogeneity and overlap. 
This work finds that the growth of infection supported by larger hyperedges can be systematically delayed until pairwise and small-group transmission have established a sufficient infected core, at which point higher-order channels become active only after a substantial transient.
These findings show that in synergistic higher-order models, the same nonlinearities that generate abrupt transitions also naturally create long subcritical phases and delayed activation of group interactions.

In competitive-inhibition contagion, inertia appears both as delayed outbreaks of a given process and as slow selection between competing endemic states. On simplicial multiplex networks, ~\cite{li2022two} investigate two competing irreversible epidemics and report that the presence of a rival contagion with strong simplex-level transmission can postpone the outbreak of the focal epidemic, effectively shifting its critical point to larger transmission rates and extending the parameter region where the infection remains rare. In a different setting, ~\cite{gracy2024competitive} analyze a competitive bivirus SIS model on hypergraphs and identify parameter domains with tri-stability, where two single-virus endemic equilibria and a coexistence equilibrium are all locally stable. Although their analysis focuses on equilibria, such a landscape implies that trajectories starting near separatrices between basins of attraction can experience prolonged mixed-infection phases before converging, so that higher-order competitive interactions act as a dynamical “brake” on the eventual dominance of any one contagion. Analogous delay effects have long been documented in pairwise multiplex competitions between epidemics and awareness processes, where the disease can only take off after an awareness process has saturated or weakened~\cite{granell2014competing}. Higher-order competition thus inherits, and often amplifies, the slow transient selection typical of multi-process spreading.

Coupled awareness–epidemic dynamics with higher-order structure add another layer to propagation delay. ~\cite{liu2023epidemic} study a two-layer system where awareness diffuses on a simplicial information layer and modifies infection probabilities on a physical contact layer. They show that simplex-enhanced awareness can raise epidemic thresholds and reshape epidemic trajectories, with parameter regimes in which infection remains suppressed for long intervals before any noticeable outbreak, or never escapes a low-prevalence plateau. Related models where awareness spreads via higher-order interactions or across multiplex structures with simplicial layers likewise report postponed epidemic peaks and extended pre-outbreak plateaus as awareness parameters or interlayer couplings increase~\cite{chang2023combined}. In these cases, path dependence arises not only from memory in the contagion itself, but also from the history of a competing information process whose higher-order pathways can stay active long after the initial epidemic shock, thereby maintaining a persistent inertial barrier to large outbreaks.

In memory-aware contagion, inertia appears because transition rates depend on past exposure or “age”. ~\cite{lin2024higher,lin2025impacts} show that in a non-Markovian social contagion model on simplicial complexes, suitable waiting-time distributions and strong higher-order connectivity keep the system in low-prevalence states for long intervals and can even change a continuous transition into a discontinuous one with hysteresis, implying long dwell times before outbreaks. ~\cite{gallo2024higher} show that empirical temporal hypergraphs exhibit long-range group-level temporal correlations and non-Markovian memory, so that effective group contact rates adjust slowly and contagion relaxes much more slowly than on memoryless temporal networks.

In threshold-activated contagion, delays originate from the need to satisfy group-level activation conditions before spreading can proceed.~\cite{mancastroppa2023hyper} reveal that hyper-core decomposition of hypergraphs reveals central higher-order cores in which nonlinear contagion localizes and from which global spreading can be efficiently seeded, implying that activity started in peripheral regions can remain trapped for long transients before reaching these cores. Bootstrap and high-order bootstrap percolation~\cite{peng2025bootstrap,wang2022generalized} on hypergraphs further show that, even when the final transition in the active set is abrupt, the number of update steps before stabilization can scale very large, revealing long precritical build-up and broad cascade lifetimes. These findings point to a structural form of inertia: higher-order cores and group thresholds systematically postpone large-scale activation even when local conditions are already favorable for spreading.

Across these mechanisms, propagation inertia emerges from insufficient early reinforcement, group-level thresholds, history dependence, or antagonistic interactions. Higher-order structures intensify these delays by requiring coordinated activation within simplices, hyperedges, or temporal groups. As a result, early-time stagnation followed by delayed cascades is a common dynamical signature of higher-order contagion, distinguishing it sharply from the immediate acceleration typical of pairwise spreading models.

\subsection{Synthesis of three-level relations}
The studies above show that similar dynamical regimes can arise from different combinations of mechanisms and higher-order structures, but with distinct structural “amplifiers” and observable signatures. Table~\ref{tab:mech_struct_behaviors} brings these observations together and highlights three recurrent patterns. First, multistability and coexistence typically arise when strong higher-order reinforcement or competition is embedded in cohesive group motifs, such as order-dependent simplex channels, nonlinear hyperedge kernels, and higher-order cores, so that multiple endemic branches coexist and become highly sensitive to initial seeding and parameter sweeps. Second, discontinuous and hybrid phase transitions are associated with sharp group-level activation rules and strongly nonlinear response functions on carefully structured simplicial and hypergraph topologies, where higher-order $k$-cores, tuned overlap, and signed or directed motifs generate abrupt jumps, hysteresis, and multicritical behavior. Third, propagation inertia and delayed outbreaks emerge when synergistic reinforcement, memory-aware contagion, and competitive–inhibition act on temporally correlated or multilayer higher-order architectures, producing long plateaus, slow attractor selection, and postponed cascades under awareness or coupling. Overall, Table~\ref{tab:mech_struct_behaviors} makes explicit that none of these behaviors is determined by structure or mechanism alone; rather, they arise from specific pairings of mechanisms with amplifying higher-order motifs.

\begin{figure}[t]
  \centering
  \includegraphics[width=\linewidth]{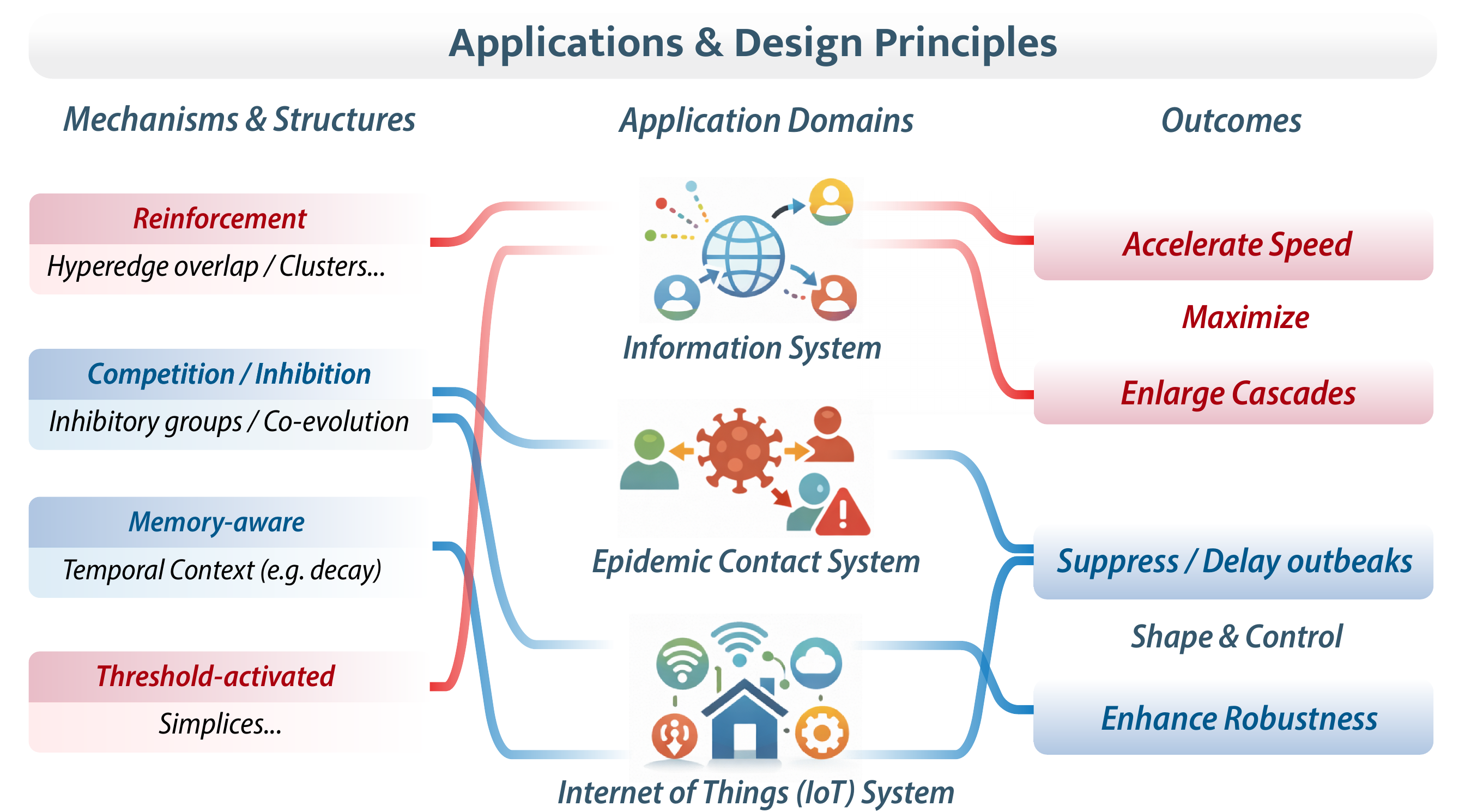}
  \caption{Examples of applications of the structure–mechanism–behavior mapping in higher-order spreading systems across information, epidemic, and IoT domains, organized into two main objectives: maximizing diffusion (e.g., accelerated speed, enlarged cascade size) and controlling spreading (e.g., suppression, temporal shaping, enhanced robustness).}
  \label{figure7}
\end{figure}

\section{Applications}\label{applications}
This section reviews representative application domains in which higher-order interactions can be leveraged to shape spreading outcomes. Rather than organizing applications purely by domain, we adopt a mechanism-oriented perspective, highlighting how spreading mechanisms interact with higher-order structures across information, epidemic, and cyber--physical systems (Fig.~\ref{figure7}) and interpreting these studies through the structure--mechanism--behavior framework.

\subsection{Maximizing diffusion via reinforcement mechanisms}
Reinforcement-driven spreading, or complex contagion, can be exploited to accelerate adoption, enlarge cascade size, and shape diffusion timelines. In social information systems, clustered networks and overlapping communities increase the effect of repeated exposure and facilitate global cascades beyond what pairwise models predict \cite{centola2010spread,ugander2012structural}. In epidemiological settings, higher-order contact structures such as households and gathering groups can amplify transmission and lead to larger or faster outbreaks \cite{landry2020effect,ma2024impact}. Similar effects also arise in IoT and cyber--physical systems, where coordinated device groups or broadcast interactions can speed up firmware dissemination and collective state alignment \cite{wan2024dynamic}.

From a design perspective, diffusion can be further enhanced by introducing group interactions, exploiting higher-order connectivity, and selecting seeds that are embedded in multiple influential groups rather than relying only on pairwise centrality \cite{iacopini2019simplicial,st2022influential,watts2002simple}. Beyond cascade size, reinforcement mechanisms also affect temporal unfolding, as synchronized or repeated exposures may induce delayed activation or abrupt transitions. Overall, these studies show that aligning higher-order structure with reinforcement mechanisms can promote faster, broader, and more robust spreading across application domains.

\subsection{Shaping and controlling spreading processes}
Higher-order interactions can also be leveraged to suppress or regulate undesirable spreading, including epidemic outbreaks, misinformation diffusion, and cascading failures. In epidemic control, groups such as households or workplaces act as key transmission units, and targeting these higher-order structures can outperform purely node-centric interventions \cite{wang2024epidemic}. In social systems, tightly connected communities can be monitored to reduce reinforcement of rumors, while in IoT and cyber--physical security, regulating group communications can help contain malware or failure cascades \cite{zhu2021malware}.

Effective control also depends on the underlying spreading mechanism. Reinforcement-driven contagion can be mitigated by limiting repeated exposure or breaking reinforcement loops, whereas memory-aware and non-Markovian processes may require temporally adaptive interventions such as quarantine, awareness, or delayed response \cite{bansal2010dynamic,granell2013dynamical}. 
Learning-based methods, including reinforcement learning, further provide adaptive ways to discover intervention policies that implicitly capture structure--mechanism coupling \cite{libin2020deep}.

Overall, existing applications suggest that higher-order spreading systems should be designed or intervened upon by jointly considering structure, mechanism, and operating regime. Features that amplify diffusion in one setting may suppress it in another, so the structure--mechanism--behavior perspective serves not only as an interpretive framework but also as a practical guide for application-specific design.

\section{Open issues}\label{openissues}
This review has approached higher-order contagion through structure–mechanism–behavior lens. 
Such a three-level mapping is particularly valuable for the design of intervention and control strategies, as it allows one to anticipate how structural or rule-based modifications translate into specific dynamical outcomes, thereby enabling targeted and efficient guidance of spreading processes.
Conversely, when emergent contagion behaviors are observed empirically, the proposed three-level correspondence provides a systematic route to identifying minimal yet adequate modeling choices, reducing unnecessary model complexity and computational cost. In this sense, the framework serves not only as a descriptive taxonomy, but also as a practical guide for the decomposition and modeling of spreading dynamics in higher-order complex systems.
In what follows, we outline several
open problems for future work in modeling, analysis, and
applications of spreading dynamics on higher-order networks.

\subsection{Structural representations}
From a structural viewpoint, several questions remain open. First, we still lack a systematic understanding of representation minimality and equivalence: when can a higher-order contact pattern be faithfully reduced to a multigraph or multilayer graph without losing the behaviors of interest, and when is an explicitly higher-order representation indispensable? Second, the role of mesoscopic features, like overlapping groups, higher-order cores and components, motifs in directed higher-order networks, remains only partially explored, especially in temporal and multiplex settings. Third, for empirical systems, we need principled ways to reconstruct and coarse-grain higher-order structure from incomplete data, and to quantify which structural aspects are dynamically relevant for a given contagion process and which can be safely ignored. Addressing these issues will be crucial for turning higher-order contagion models into reliable tools rather than purely theoretical constructs.

\subsection{Dynamical regimes and theoretical challenges}
At the behavioral level, higher-order contagion exhibits a striking set of phenomena that are rare or absent in purely pairwise models. Yet our theoretical understanding of these regimes is still fragmentary. Existing analyzes often focus on specific models on specific substrates, making it difficult to identify robust mechanisms and universality classes across formulations.

Open problems span several directions. We lack a systematic bifurcation and stability theory for higher-order contagion that can organize the observed multistable and hybrid transitions in a common language, and that can incorporate the effects of finite-size fluctuations, structural heterogeneity, and temporal variability. The interplay between the three regimes, how multistability, discontinuous thresholds, and inertia co-occur or exclude each other across parameter space, also remains poorly charted. Finally, little is known about the control and optimization of these behaviors: for example, how to exploit higher-order structure to suppress undesirable discontinuous outbreaks, or conversely to stabilize coexistence in multi-strain or multi-opinion systems.

\subsection{Data, inference, and applications}
A final set of challenges concerns data and applications. Most empirical work to date uses higher-order structures extracted from co-occurrence data, temporal contact traces, or interaction logs, but we still lack standard pipelines for inferring higher-order contagion mechanisms from time series and for performing model selection across competing structure–mechanism combinations. Developing statistical and machine-learning tools that can distinguish, for example, synergistic contagion from threshold-activated or memory-aware dynamics based on observational data, and that can learn effective infection kernels or threshold functions from data, would greatly increase the practical value of higher-order models.

On the application side, higher-order contagion has clear relevance for problems such as epidemic control in clustered settings, coordination and polarization in social systems, and cascading failures in engineered infrastructure. Turning qualitative insights into actionable design principles, for instance, which group structures to avoid to reduce the risk of discontinuous outbreaks, or how to exploit awareness layers to induce benign forms of coexistence, is still an open question. Bridging the gap between theory and practice will likely require closer interaction between network scientists, domain experts and data-driven modelling communities, as well as tighter integration with developments in graph and hypergraph learning, intervention optimisation and early-warning methods.

\section{Conclusion}\label{conclusion}

In this survey, we have reviewed spreading dynamics on higher-order networks from a structure--mechanism--behavior perspective. 
Moving beyond model-centric descriptions, we emphasized how different higher-order representations shape interaction patterns, how spreading mechanisms operate on these structures, and how their interplay gives rise to diverse emergent behaviors.

Building on this perspective, we provided an overview that connects modeling, analysis, and control of spreading processes. 
This mapping not only clarifies the relationships among existing models, but also offers practical guidance for designing maximization and intervention strategies, from structure-aware modeling to mechanism-driven control.




\ifCLASSOPTIONcaptionsoff
  \newpage
\fi



%
\bibliographystyle{IEEEtran}
\bibliography{bibliography.bib}

\begin{IEEEbiography}[{\includegraphics[width=1in,height=1.25in,clip,keepaspectratio]{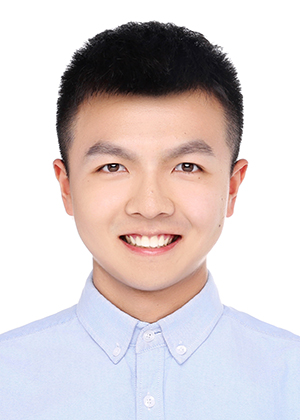}}]{Ming Xie} (Student Member, IEEE) received the M.S. degree in electronic information from Alibaba Complex Science Research Center, Hangzhou Normal University, Hangzhou, China. He is currently working toward the Ph.D. degree in control science and engineering with the College of Control Science and Engineering, Zhejiang University, Hangzhou, China, supervised by Prof. Shibo He and Prof. Jiming Chen. 

His research interests include network science, higher-order networks, and machine learning.
\end{IEEEbiography}

\begin{IEEEbiography}[{\includegraphics[width=1in,height=1.25in,clip,keepaspectratio]{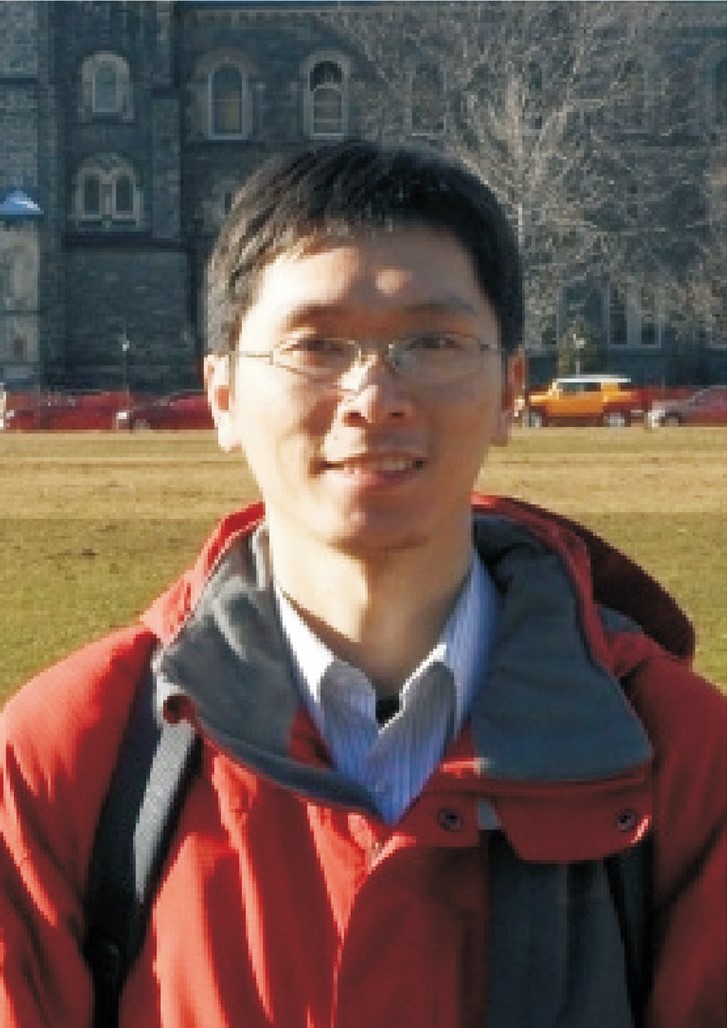}}]{Shibo He} (Senior Member, IEEE) received the Ph.D. degree in control science and engineering from Zhejiang University, Hangzhou, China, in 2012.

He is currently a Professor with Zhejiang University. His research interests include Internet of Things, crowdsensing, and big data.

Dr. He serves on the editorial board of several journals. He served as a Symposium Co-Chair for IEEE ICC 2017, the Finance Registration Chair of ACM MobiHoc 2015, a Technical Program Committee Co-Chair for IEEE ScalCom 2014, etc.
\end{IEEEbiography}

\begin{IEEEbiography}
[{\includegraphics[width=1in,height=1.25in,clip,keepaspectratio]{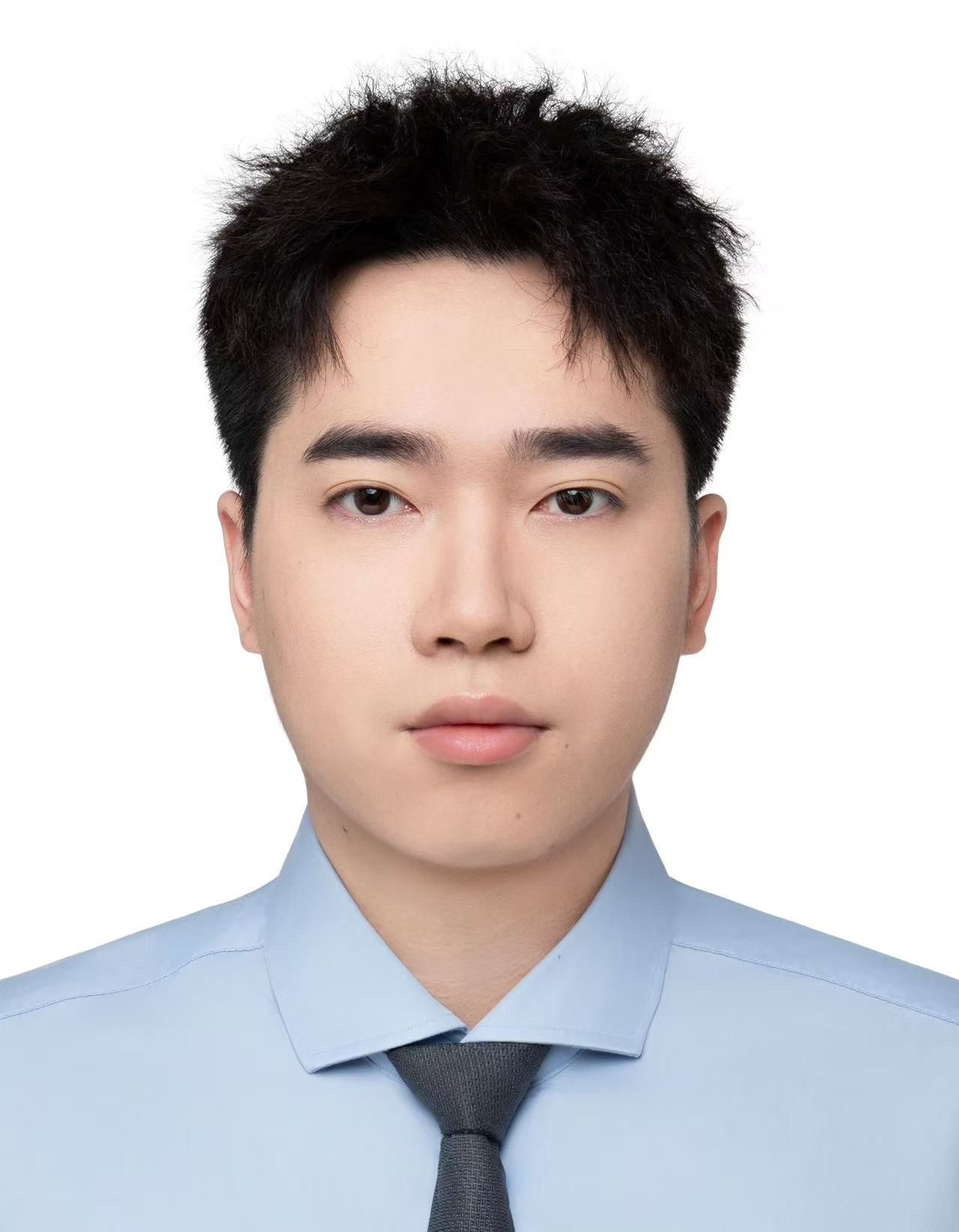}}]
{Qihao Huang} He is currently working toward the Ph.D. degree in control science and engineering with the College of Control Science and Engineering, Zhejiang University, Hangzhou, China, supervised by Prof. Shibo He and Prof. Jiming Chen. 
\end{IEEEbiography}

\begin{IEEEbiography}[{\includegraphics[width=1in,height=1.25in,clip,keepaspectratio]{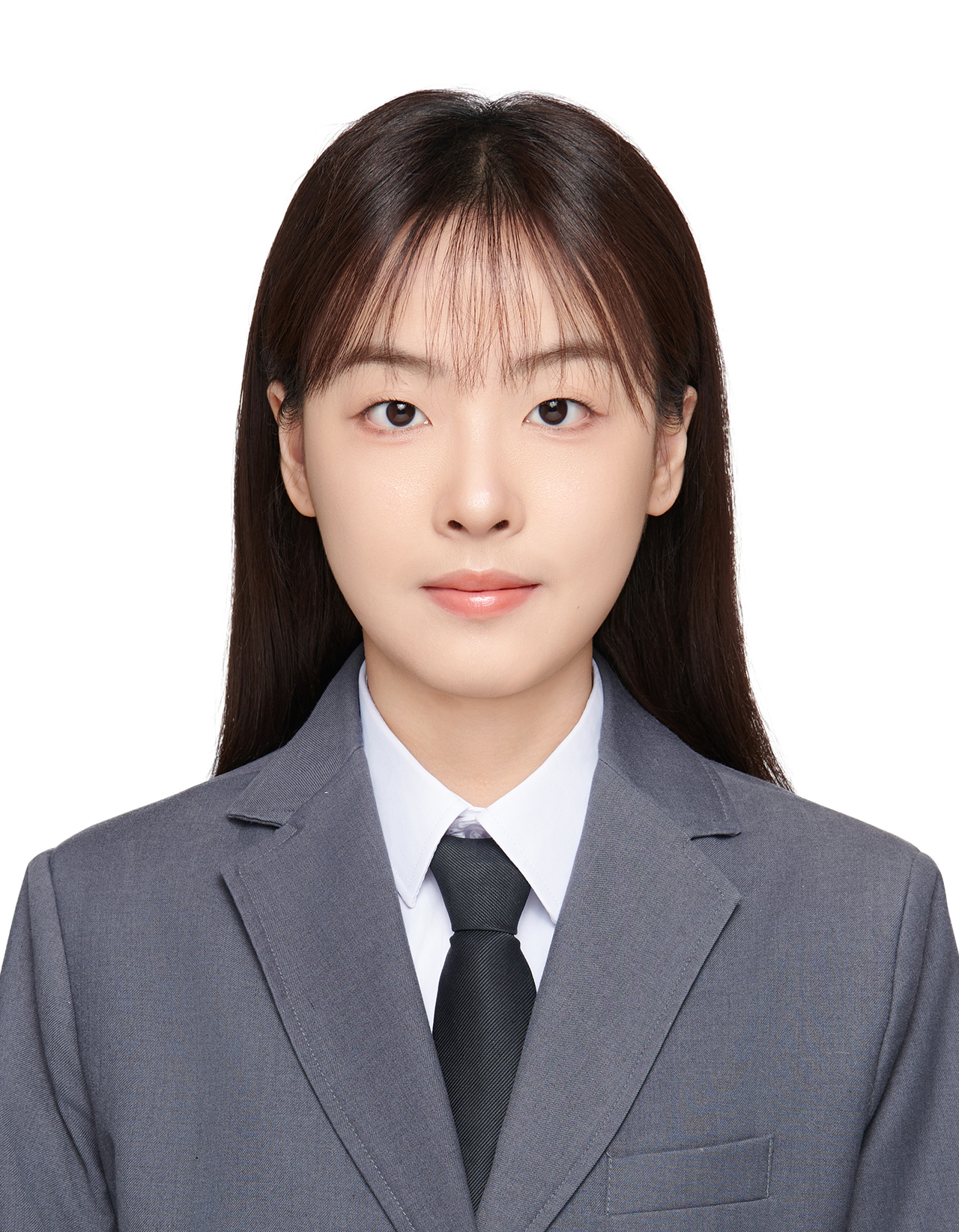}}]{Yuzhen Fan}
received the B.Sc. degree in Information and Computational Science from the College of Science in Northeastern University, China in 2020, and the Ph.D. degree from the School of Mathematical Sciences, Zhejiang University, China in 2025. She is currently a Postdoctoral researcher in the College of Control Science and Engineering at Zhejiang University.

Her research interests are in the areas of control theory, synthetic biology, and chemical reaction networks theory.
\end{IEEEbiography}

\begin{IEEEbiography}
[{\includegraphics[width=1in,height=1.25in,clip,keepaspectratio]{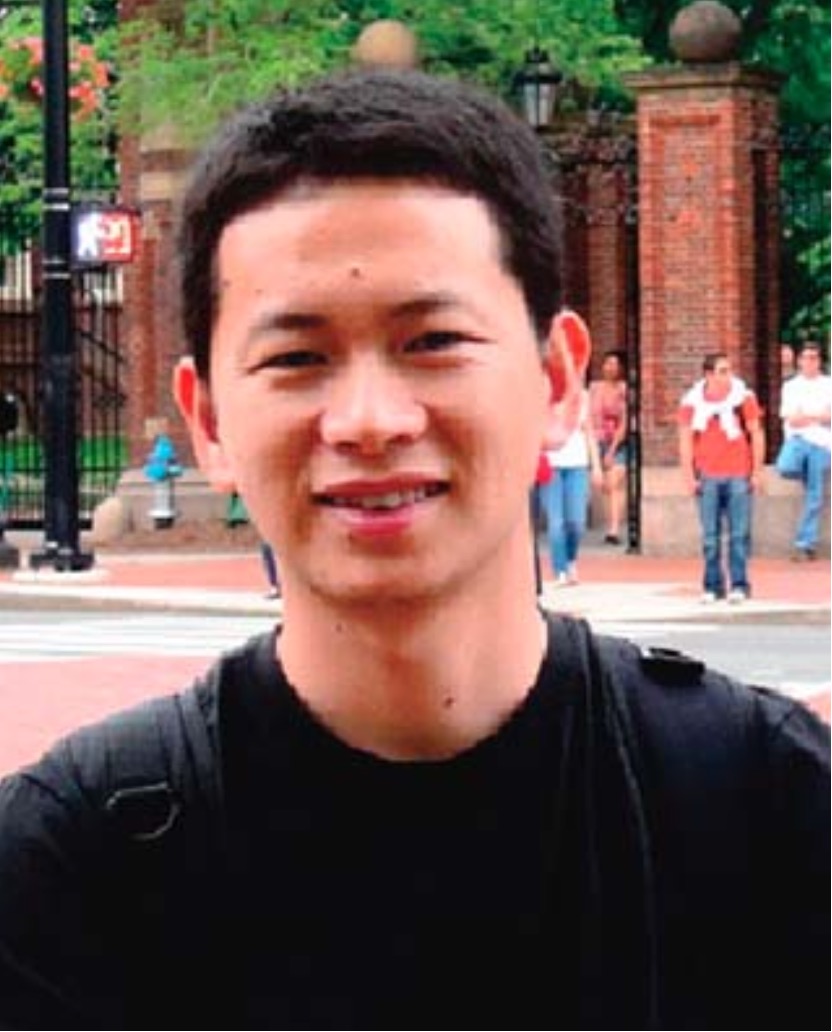}}]
{Jiming Chen} (Fellow, IEEE) received the Ph.D. degree in control science and engineering from Zhejiang University, Hangzhou, China, in 2005.

He is currently a Professor with the Department of Control Science and Engineering, Zhejiang University. His research interests include Internet of Things, networked control, and wireless networks.

Dr. Chen serves on the editorial boards of multiple IEEE Transactions, and a General Co-Chair for IEEE RTCSA’19, IEEE Datacom’19, and IEEE PST’20. He was a recipient of the 7th IEEE ComSoc Asia/Pacific Outstanding Paper Award, the Japan Society for the Promotion of Science Invitation Fellowship, and the IEEE ComSoc AP Outstanding Young Researcher Award. He is the Distinguished Lecturer of IEEE Vehicular Technology Society distinguished lecturer. He is a fellow of the CAA.
\end{IEEEbiography}

\end{document}